\documentclass[10pt,secnumarabic,nofootinbib,aps,twocolumn,superscriptaddress,preprintnumbers,prd]{revtex4-2}
\pdfoutput=1
\usepackage[utf8]{inputenc}
\usepackage{color}
\usepackage{babel}
\usepackage{refstyle}
\usepackage{mathtools}
\usepackage{amsmath}
\usepackage{amsthm}
\usepackage{amssymb}
\usepackage{graphicx}
\usepackage{setspace}
\usepackage{microtype}

\makeatletter

\AtBeginDocument{\providecommand\Figref[1]{\ref{Fig:#1}}}
\RS@ifundefined{subsecref}
  {\newref{subsec}{name = \RSsectxt}}
  {}
\RS@ifundefined{thmref}
  {\def\RSthmtxt{theorem~}\newref{thm}{name = \RSthmtxt}}
  {}
\RS@ifundefined{lemref}
  {\def\RSlemtxt{lemma~}\newref{lem}{name = \RSlemtxt}}
  {}

\usepackage{tikz}
\usetikzlibrary{backgrounds}
\usepackage{slashed}	 	
\usepackage{amsfonts} 		
\SetTracking{encoding={*}, shape=sc}{0} 	
\usepackage{color} 		
\definecolor{darkblue}{rgb}{0.0,0.0,0.4} 	
\definecolor{darkred}{rgb}{0.4,0.0,0.0} 	
\usepackage[colorlinks]{hyperref}
\hypersetup{pdftitle={Machine Learning Line Bundle Connections},
	pdfauthor={},
	linkcolor=darkblue, urlcolor=darkblue, citecolor=darkblue, filecolor=darkblue, linktoc=page}
	
\allowdisplaybreaks		

\newcommand{\newsection}[1]{\noindent \textbf{#1}.}

\makeatletter
\g@addto@macro\bfseries{\boldmath}
\makeatother

\makeatletter
\g@addto@macro\@floatboxreset\centering
\makeatother

\let\originalleft\left
\let\originalright\right
\renewcommand{\left}{\mathopen{}\mathclose\bgroup\originalleft}
\renewcommand{\right}{\aftergroup\egroup\originalright}

\usepackage{tikz} 
\usetikzlibrary{shapes.geometric,arrows} 

\tikzstyle{block} = [rectangle, draw,    
text width=15em, text centered, inner ysep = 1em, minimum height=1em] 
\tikzstyle{netblock} = [rectangle, draw, color=darkblue,   
text width=10em, text centered, inner ysep = 1em, minimum height=1em]
\tikzstyle{netblock2} = [rectangle, draw, color=darkred,  rounded corners,   
text width=10em, text centered, inner ysep = 1em, minimum height=1em ] 
\tikzstyle{netblock3} = [rectangle, draw, color=darkblue,   
text width=8em, text centered, inner ysep = 1em, minimum height=1em ]
\tikzstyle{netblock4} = [rectangle, draw, color=darkblue,   
text width=3.5em, text centered, inner ysep = 0.0em, minimum height=1.5em ]
\tikzstyle{line} = [draw, -latex']

\makeatother

\global\long\def\SU#1{\text{SU}(#1)}%
\global\long\def\vol{\operatorname{vol}}%
\global\long\def\Vol{\operatorname{Vol}}%
\global\long\def\tr{\operatorname{tr}}%
\global\long\def\op#1{\operatorname{#1}}%
\global\long\def\eqspace{\mathrel{\phantom{{=}}{}}}%
\global\long\def\bP{\mathbb{P}}%
\global\long\def\id{\operatorname{id}}%
\global\long\def\id{\boldsymbol{1}}%

\begin{document}

\preprint{LIMS-2021-13}

\title{Machine Learning Line Bundle Connections}
\author{Anthony Ashmore}
\email[]{ashmore@uchicago.edu}
\affiliation{Kadanoff Center for Theoretical Physics, University of Chicago,, IL 60637, USA}
\affiliation{Sorbonne Universit\'e, CNRS, Laboratoire de Physique Th\'eorique et Hautes Energies, F-75005 Paris, France}
\author{Rehan Deen}
\email[]{rehan.deen@gmail.com}
\affiliation{Rudolf Peierls Centre for Theoretical Physics, University of Oxford, OX1 3PU, UK}
\author{Yang-Hui He}
\email[]{hey@maths.ox.ac.uk}
\affiliation{London Institute for Mathematical Sciences, Royal Institution, W1S 4BS, UK}
\affiliation{Department of Mathematics, City, University of London, EC1V0HB, UK}
\affiliation{Merton College, University of Oxford, OX1 4JD, UK}
\affiliation{School of Physics, NanKai University, Tianjin, 300071, P.R.~China}
\author{Burt A.\,Ovrut}
\email[]{ovrut@elcapitan.hep.upenn.edu}
\affiliation{Department of Physics, University of Pennsylvania, Philadelphia, PA 19104, USA}

\begin{abstract}
\noindent We study the use of machine learning for finding numerical hermitian Yang--Mills connections on line bundles over Calabi--Yau manifolds. Defining an appropriate loss function and focusing on the examples of an elliptic curve, a K3 surface and a quintic threefold, we show that neural networks can be trained to give a close approximation to hermitian Yang--Mills connections.
\end{abstract}

\maketitle

\section{Introduction and Summary}

\noindent Heterotic string theory on Calabi--Yau threefolds equipped with gauge bundles provide a large class of phenomenologically promising string models~\cite{Braun:2005ux,Lukas:1998yy,Donagi:1999ez,Bouchard:2005ag,Blumenhagen:2006ux,Lebedev:2006kn,Candelas:2007ac,Lebedev:2008un,MayorgaPena:2012ifg,Anderson:2009mh,Anderson:2011ns,Anderson:2013xka}. However, despite many decades of work, it is still not possible to compute the masses or couplings that appear in the resulting four-dimensional theories from first principles. A good deal of the problem can be traced to the lack of explicit expressions for non-trivial Calabi--Yau (CY) metrics or hermitian Yang--Mills connections. Let us recall why these are needed. Compactification of the heterotic string on a Calabi--Yau threefold $X$ with its Ricci-flat metric gives a four-dimensional effective theory with $N=1$ supersymmetry. To obtain MSSM-like theories, $X$ should also carry a holomorphic vector bundle $V$ whose connection solves the hermitian Yang--Mills (HYM) equations~\cite{Donaldson, UhlenbeckYau}. 

Generic details of the compactification, such as the number of generations or the vanishing of certain couplings, can be obtained from algebro-geometric results for the existence and topology of the threefold $X$ and the bundle $V$~\cite{Greene:1986ar,Greene:1986bm,Greene:1986jb,Matsuoka:1986vg,Greene:1987xh,Donagi:2000zs,Braun:2006me,Anderson:2010tc}. These calculations do not need explicit expressions for either the metric or the connection. Unfortunately, the detailed four-dimensional physics is controlled by a K\"ahler potential and a superpotential, which depend on both the explicit Ricci-flat metric and the explicit HYM connection. Without this data, it is generally not possible to accurately compute masses or couplings, leaving us unable to make precise particle physics predictions from string theory.

With little hope of finding analytic expressions for the relevant metrics or connections, much progress has been made on finding \emph{numerical} approximations. There is now a diverse range of algorithms for computing Ricci-flat metrics on Calabi--Yau manifolds numerically, including position space methods~\cite{Headrick:2005ch}, spectral approaches~\cite{Douglas:2006rr,Braun:2007sn,Headrick:2009jz} building on the work of Tian~\cite{Tian} and Donaldson~\cite{math/0512625}, and, most recently machine learning~\cite{Ashmore:2019wzb} and neural networks tailored for the metric computation \cite{Anderson:2020hux,Douglas:2020hpv,Jejjala:2020wcc,Douglas:2021zdn} (see \cite{He:2018jtw} for a recent pedagogical review on Calabi--Yau manifolds and machine-learning).

Given these advances, it now seems appropriate to focus on computing hermitian Yang--Mills connections. This will be the subject of the present work. As with the Ricci-flatness condition for the metric, the HYM equations are a system of partial differential equations that are difficult to solve, and so one is again compelled to consider numerical approximations. Previous work~\cite{Douglas:2006hz,Anderson:2010ke,Anderson:2011ed} has used Wang's extension~\cite{Wang} of Donaldson's approach to compute numerical HYM connections for a number of examples, including $\SU n$ bundles over threefolds.

Unfortunately, connections are a jump in computational complexity compared to the Ricci-flat metric, with a corresponding loss of speed and accuracy. With this in mind, we seek a faster and more accurate method by employing machine learning. The aim of this paper is to take the first step in applying machine learning to find HYM connections. Focusing on the simplest cases of connections on line bundles, we show that it is both feasible and promising to compute connections in this way. Though we do not tackle non-abelian bundles in the present work, we note that many Standard Model-like theories can be obtained from heterotic line bundle models~\cite{Anderson:2009mh,Anderson:2011ns,Anderson:2012yf,Anderson:2013xka,GrootNibbelink:2015lme,GrootNibbelink:2015dvi,GrootNibbelin:2016ovb,Braun:2017feb}

Our approach builds on and extends the work of Douglas et al.~\cite{Douglas:2020hpv} which presented a neural network for computing Calabi--Yau metrics (they have provided a TensorFlow implementation of their approach on GitHub~\cite{MLGeometry}). We give three examples, namely line bundles over an elliptic curve, a K3 surface, and a quintic threefold. For each of these, one starts by computing a numerical approximation to the Calabi--Yau metric. One then constructs a neural network whose input is the coordinates on the Calabi--Yau and whose output is interpreted as the hermitian metric on the line bundle. By taking derivatives of the neural network, one can compute both the connection and the curvature defined by the hermitian metric. We give a loss function whose value is minimised for hermitian Yang--Mills connections, and then use this loss function to train the network. The resulting network encodes the hermitian metric that defines a HYM connection on the Calabi--Yau. In this way, we find that accurate HYM connections can be obtained in a straightforward manner. In our results, we examine how the accuracy of the numerical connections changes with varying network depth. We observe that deeper networks are generally more accurate (as expected since they contain more parameters), with this improvement more pronounced as the dimension of the Calabi--Yau increases.

There are a number of obvious extensions. First, one could consider line bundles over more complicated Calabi--Yau manifolds. This would involve generalising the code of Douglas et al.~\cite{Douglas:2020hpv} to complete intersections in products of projective space. Second, one would want to move beyond abelian bundles to consider non-abelian bundles, defined by monads, extensions, and so on. Both of these are essential if one wants to make contact with the many constructions of the so-called heterotic Standard Model~\cite{hep-th/0512177,hep-th/0502155,hep-th/0501070,hep-th/0512149,0911.1569,1112.1097,1106.4804,1202.1757,1307.4787,1506.00879,1507.07559,1007.0203,hep-th/9903052}. These advances, together with numerical metrics and Laplacians~\cite{Braun:2008jp,Ashmore:2020ujw,Afkhami-Jeddi:2021qkf}, and results for the matter-field K\"ahler potential~\cite{hep-th/9902071,McOrist:2016cfl,1801.09645,Ishiguro:2021drk}, should enable real progress on computing masses and couplings in top-down string models. We will discuss these issues in future publications.


\section{Hermitian Yang--Mills and line bundles on Calabi--Yau manifolds\label{sec:Line-bundle-backgrounds}}

\noindent Given a complex manifold $X$ with a  K\"ahler metric $g$ (defined by a choice of complex structure and  K\"ahler form $J$), a stable holomorphic vector bundle $V$ admits a unique connection $A$ whose curvature $F$ solves the hermitian Yang--Mills equations:
\begin{equation}
F_{ij}=F_{\bar{i}\bar{j}}=0,\qquad g^{i\bar{j}}F_{i\bar{j}}=\mu(V)\,\id.\label{eq:HYM}
\end{equation}
Here $g^{i\bar{j}}$ is the inverse  K\"ahler metric on $X$, $\mu(V)$ is a real constant known as the \emph{slope} of $V$, and $\id$ is the $d\times d$ identity matrix on the fibres of the rank-$d$ bundle $V$. The first two conditions are equivalent to the holomorphicity of $V$ (and will be automatic in our construction). The third condition gives the HYM equations, a system of non-linear PDEs for the connection $A$. 

The connection $A$ can equivalently be described by a hermitian structure on $V$, which, more prosaically, is simply a hermitian inner product $G$ on sections of $V$. Given a frame $\{e_{a}\}$ for $V$, the inner product is
\begin{equation}
(e_{a},e_{b})=G_{\bar{a}b},\qquad G=G^{\dagger}.
\end{equation}
In holomorphic gauge, the connection is determined by $G$ as
\begin{equation}
A_{i}=G^{-1}\partial_{i}G,\qquad A_{\bar{i}}=0,
\end{equation}
with the curvature then given by
\begin{equation}
F_{i\bar{j}}=\partial_{\bar{j}}\partial_{i}\log G,
\end{equation}
where we are using the shorthand notation $\partial_{i}\log G\equiv G^{-1}\partial_{i}G$. Given a  K\"ahler metric, finding a solution to the HYM equations then reduces to choosing $G$ such that (\ref{eq:HYM}) is satisfied. If this is the case, $G$ is known as a \emph{Hermite--Einstein} metric on $V$.

There exists a solution to the HYM equations on a  K\"ahler manifold if and only if the holomorphic vector bundle $V$ is (at least) polystable~\cite{Donaldson,UhlenbeckYau}. To check this, one begins by computing the slope of $V$ via
\begin{equation}
\mu(V)\equiv\int_{X}c_{1}(V)\wedge J^{n-1}.
\end{equation}
Note that we always normalise $\Vol_{g}$, the volume of $X$ as measured by the  K\"ahler metric, to one. The bundle $V$ is stable if $\mu(\mathcal{F})<\mu(V)$ for all subsheaves $\mathcal{F}\subset V$ with $0<\op{rank}\mathcal{F}<\op{rank}V$. Polystability is the statement that $V$ is a direct sum of stable bundles, all with the same slope. Thanks to this, the \emph{existence} of a HYM connection can be reduced to algebraic conditions on subsheaves of $V$. Notice however that this is in no way constructive; that is, knowing a HYM connection exists does not give any hint of how to find it explicitly. For this we must turn to numerical methods. The aim of the present work is to use a neural network to search for numerical HYM solutions for the simplest examples, namely line bundles.

Line bundles on CY manifolds are by now a well-understood ingredient in heterotic compactifications (see, for example, \cite{Anderson:2009mh,Anderson:2011ns,Anderson:2013xka,Anderson:2012yf} and references therein). 
Recall that a holomorphic line bundle $L$ over a complex manifold is determined (up to torsion) by its first Chern class, $c_{1}(L)$. Thanks to this, we can associate a line bundle over $X$ to a divisor $\mathcal D$ by taking
\begin{equation}
c_{1}(L)\equiv\frac{[F]}{2\pi}=\mathcal D,
\end{equation}
where $[F]$ is the class of the curvature of the connection on $L$. The corresponding line bundle is then denoted by $\mathcal{O}_{X}(\mathcal D)$, or often by $\mathcal{O}_{X}(k^{I})$, where $\mathcal D=k^I \mathcal D_I$ and the basis of divisors is implicit. The slope of a line bundle is then
\begin{equation}
\mu(L)=\int_{X}c_{1}(L)\wedge J^{n-1},
\end{equation}
which depends on both the choice of line bundle via $c^1(L)$ and the choice of K\"ahler moduli via $J$. Since a line bundle has no subsheaves $\mathcal{F}\subset L$ with $0<\op{rank}\mathcal{F}<1$, line bundles are always stable. This means that a line bundle will always admit a connection that solves the HYM equation, $g^{i\bar{j}}F_{i\bar{j}}=\mu(L)$. The problem is finding the explicit form of this connection.

\section{Numerical metrics and connections from neural networks\label{sec:Numerical-metrics-and}}

\noindent In this section, we begin by reviewing the calculation of numerical Calabi--Yau metrics using a neural network following Douglas et al.~\cite{Douglas:2020hpv}.\footnote{See also \cite{Douglas:2021zdn} for a discussion of the ``holomorphic feedforward networks'' that underlie this approach.} We then discuss numerical HYM connections. Finally, we define a functional which is minimised on HYM connections and thus can act as a loss function for a suitable neural network whose output will be interpreted as $\log G^{-1}$.

\subsection{Numerical metrics from neural networks}

\noindent Consider a compact Calabi--Yau $n$-fold $X$ defined as a hypersurface in $\bP^{n+1}$ by the vanishing of a holomorphic equation $f(z)=0$ of degree $n+2$ (for example, a threefold defined by a quintic equation in $\bP^{4}$). The choice of defining equation $f$ fixes the complex structure moduli of the Calabi--Yau. A choice of K\"ahler  structure then determines the metric $g_{i\bar{j}}$ on $X$. In other words, the metric on $X$ is fixed by a K\"ahler  potential $K(z,\bar{z})$. Finding the Ricci-flat metric on $X$ then amounts to choosing $K$ such that the resulting K\"ahler  metric is Ricci flat. 

Apart from on the torus, there are no explicitly known K\"ahler  potentials that give such Ricci-flat metrics. Instead, work has mostly focused on finding numerical approximations starting from a Fubini--Study-like ansatz for $K$:
\begin{equation}
K=\frac{1}{k\pi}\log s_{\alpha}h^{\alpha\bar{\beta}}\bar{s}_{\bar{\beta}},\label{eq:FS_ansatz}
\end{equation}
where the $s_{\alpha}$ are sections of $\mathcal{O}_{X}(k)$ (homogeneous functions of the coordinates $z$ of degree $k$ modulo $f=0$), and $h^{\alpha\bar{\beta}}$ is a hermitian matrix of parameters. One then varies the parameters so that the resulting K\"ahler  metric is as close as possible to Ricci flat. Increasing the degree $k$ increases the size of the matrix $h^{\alpha\bar{\beta}}$, allowing a better approximation of the honest Calabi--Yau metric. There are now a variety of schemes for choosing $h^{\alpha\bar{\beta}}$, including via balanced metrics~\cite{math/0512625,Douglas:2006rr,Braun:2007sn}, direct optimisation~\cite{Headrick:2005ch,Headrick:2009jz} and neural networks~\cite{Anderson:2020hux}.

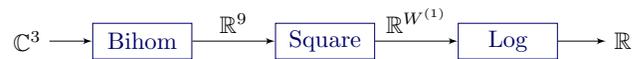
\begin{figure}[t]
   \scalebox{0.9}{
\begin{tikzpicture}[node distance = 2.0cm, auto]
    \node  (b0) {$\mathbb{C}^3$};
    \node [netblock4, right of = b0, xshift=-0.3cm] (b1) {Bihom};
    \node [netblock4, right of = b1, xshift=0.7cm](b2) { Square };
    \node [netblock4, right of = b2, xshift=0.7cm](b3) { Log };
    \node [right of = b3, xshift=-0.3cm](b4) { $\mathbb{R}$};
    \path [line] (b0) -- (b1) ;
    \path [line] (b1) -- node[above] {$\mathbb{R}^{9}$} (b2);
    \path [line] (b2) -- node[above] {$\mathbb{R}^{W^{(1)}}$} (b3);
    \path [line] (b3) -- (b4);
  \end{tikzpicture}
}

\caption{A $D=2$ network on an elliptic curve, whose output should be interpreted either as the K\"ahler potential, $K$, or log of the inverse bundle metric, $\log G^{-1}$, depending on whether one is computing the Calabi--Yau metric or the hermitian Yang--Mills connection. Here, ``Bihom'' refers to a bihomogenous layer which takes $z_i=(z_0,z_1,z_2)$ as input and outputs the real and imaginary parts of $z_i \bar{z}_{\bar{j}}$. ``Square'' is a dense layer with a quadratic activation function, $\vec x \mapsto (W_1 \vec x)^2$, where $W_1$ is a general linear transformation of dimension $W^{(1)} \times 9$. ``Log'' is a dense layer with a $\log$ activation function, $\vec x \mapsto \log(W_2 \vec x)$, where $W_2$ is a general linear transformation of dimension $1\times W^{(1)}$.}
\label{fig:linear_network}
\end{figure}

The work of Douglas et al.~\cite{Douglas:2020hpv} follows a similar path but uses a neural network to compute the K\"ahler potential directly (see also \cite{Anderson:2020hux,Jejjala:2020wcc} for similar approaches). The network is a series of densely connected layers $L^{(i)}$ of depth $D$ and width $W^{(i)}$ with quadratic activation functions $\theta^{(i)}\colon x\mapsto x^{2}$. The final layer, $L^{(D)}$, has width $W^{(D)}=1$ and a $\log$ activation function, $\theta^{(D)}\colon x\mapsto\log x$. The output of the network can thus be thought of as the logarithm of a homogeneous scalar function of the inputs, with the coefficients that appear in this function fixed by the collective weights $\boldsymbol{v}$ of the network. A diagram of this network structure is shown in Figure \ref{fig:linear_network}.

The inputs to the network are coordinates on the Calabi--Yau hypersurface, given as points $z_i=[z_{0}:\ldots:z_{n+1}]$ in the ambient projective space $\bP^{n+1}$, which can thus be thought of as sections of $\mathcal{O}_{X}(1)$, i.e.~elements of $H^{0}(X,\mathcal{O}_{X}(1))$. In practice, the first layer is actually a ``bihomogeneous layer'' which converts the inputs $z_i$ to the real and imaginary parts of $z_i \bar{z}_{\bar j}$, allowing one to work with real quantities. The successive layers have activation functions which square the output of each layer, so that the network essentially constructs the tensor product
\begin{equation}
\bigotimes_{1}^{D-1}\mathcal{O}_{X}(2)=\mathcal{O}_{X}(2^{D-1}).\label{eq:layers}
\end{equation}
Thus the output of the penultimate layer represents elements of $H^{0}(X,\mathcal{O}_{X}(2^{D-1}))$. Together with the final layer, the network output is $K$, the K\"ahler  potential, with the precise way that elements of $H^{0}(X,\mathcal{O}_{X}(2^{D-1}))$ are combined fixed by the weights $\boldsymbol{v}$. The output of the network can then be used to compute a K\"ahler metric on the hypersurface. The aim is then to choose the weights $\boldsymbol{v}$ so that the resulting metric is as close as possible to Ricci flat.

The network is trained by minimising the pointwise difference between the volume defined by the (explicitly known) holomorphic $(n,0)$-form, $\vol_{\Omega}$, and the volume defined by the K\"ahler metric on $X$, $\vol_{g}$ (using $K$ computed by the network). The two quantities agree only when the metric is the honest Ricci-flat metric. 

In outline, training proceeds as follows. First, a training set and a test set, each containing 10,000 points lying on the Calabi--Yau hypersurface, are generated. The training points (and data about coordinate patches, the $(n,0)$-form and the point distribution) are passed to the network in batches of 1,000 in a training round. The K\"ahler metric defined by the network is given by the complex Hessian of the network output, $g_{i\bar{j}}(\boldsymbol{v})\sim\partial_{i}\partial_{\bar{j}}K(\boldsymbol{v})$, where $\boldsymbol{v}$ denotes the weights of the network. The loss function is simply the mean absolute percentage error (MAPE), summed over the points in the training round:
\begin{equation}
\sigma(\boldsymbol{v})=\int_{X}\left|1-\frac{\vol_{g}(\boldsymbol{v})}{\vol_{\Omega}}\right|\vol_{\Omega},
\end{equation}
where here, and in what follows, we normalise the integrated volumes, $\Vol_{g}$ and $\Vol_{\Omega}$, to one. Note that this is known as the ``$\sigma$ measure'' in \cite{Braun:2007sn} and later work. One then searches for the minimum of this function in weight space, using stochastic gradient descent to update the weights after each training round. After 500 epochs, the network has usually converged to an approximately Ricci-flat K\"ahler potential. The accuracy of the resulting network can then be checked by evaluating $\sigma(\boldsymbol v)$ on the test set.

It is simple to see how the numerical accuracy of the approximation can be increased. From (\ref{eq:layers}), a deeper network provides a higher-degree expansion of the K\"ahler potential with more parameters (weights), both of which should allow a better approximation of the Ricci-flat metric. A wider network increases only the number of parameters (weights).

\subsection{Numerical connections from neural networks}

\noindent Following \cite{Douglas:2006hz,Anderson:2010ke,Anderson:2011ed}, one can calculate numerical HYM connections by starting with an ansatz similar in spirit to (\ref{eq:FS_ansatz}) but now for the hermitian structure $G$ as
\begin{equation}
(G^{-1})^{a\bar{b}}=\sum_{\alpha,\beta}^{N_{k}}S_{\alpha}^{a}H^{\alpha\bar{\beta}}\bar{S}_{\bar{\beta}}^{\bar{b}},
\end{equation}
where $S_{\alpha}^{a}$ are sections of $V\otimes\mathcal{O}_{X}(k)$ and $H^{\alpha\bar{\beta}}$ is a hermitian matrix of parameters. In principle, one then varies these parameters to find an approximate solution to the HYM equation (\ref{eq:HYM}). The result of this is the hermitian metric, and hence connection, on the bundle $V(k)\equiv V\otimes\mathcal{O}_{X}(k)$. Since we are interested in the connection on $V$ alone, one should subtract the contribution of $\mathcal{O}_{X}(k)$. As discussed in \cite{Anderson:2010ke,Anderson:2011ed}, the optimal way to do this is to take the metric on $\mathcal{O}_{X}(k)$ to be that induced by $\det G$. For the case where $V$ is a line bundle, one does not encounter this complication as the connection on $V=\mathcal{O}_{X}(m)$ is simple to recover from the connection on $\mathcal{O}_{X}(m+k)$. As with the metric, increasing $k$ increases the number of sections $S_{\alpha}^{a}$ and hence the number of parameters in $H^{\alpha\bar{\beta}}$, so that larger values of $k$ allow for a better approximation to the honest HYM connection.

Our idea is to use the structure of a neural network to mimic the construction of $G^{-1}$, using the coordinates and activation functions to reproduce the sections, with the weights standing in for the parameters.

We will focus on the example of rank-one bundles, i.e.~line bundles. In this case, the hermitian fibre metric on the line bundle is a scalar
\begin{equation}
G^{-1}=\sum_{\alpha,\beta}^{N_{k}}S_{\alpha}H^{\alpha\bar{\beta}}\bar{S}_{\bar{\beta}},
\end{equation}
with the curvature given by
\begin{equation}
F_{i\bar{j}}=\partial_{\bar{j}}\partial_{i}\log G=-\partial_{\bar{j}}\partial_{i}\log G^{-1}.
\end{equation}
We will treat the output of the neural network as $\log G^{-1}$, from which it is simple to calculate $F_{i\bar{j}}$.

The structure of the connection network is the same as that of \Figref{linear_network}, with the depth of the network controlling the value of $m+k$. As with the metric network, the inputs are the points on the Calabi--Yau hypersurface, given as points on $\bP^{n+1}$, and the output of the network should be identified with $\log G^{-1}$. For each training round, one computes $F_{i\bar{j}}$ as the complex Hessian of the output of the network. One then uses a previously trained metric network to compute the Ricci-flat metric, and combines this with $F_{i\bar{j}}$ into an appropriate loss function, which we give in the next subsection. Training then attempts to minimise this loss function to find a numerical approximation to the HYM connection (for a given choice of Ricci-flat metric). After sufficient training rounds, one has a neural network that is equivalent to $\log G^{-1}$ as a function of coordinates. A schematic of this structure is given in the appendices in \Figref{connection_network}.

\begin{figure*}
\includegraphics[width=0.3\textwidth]{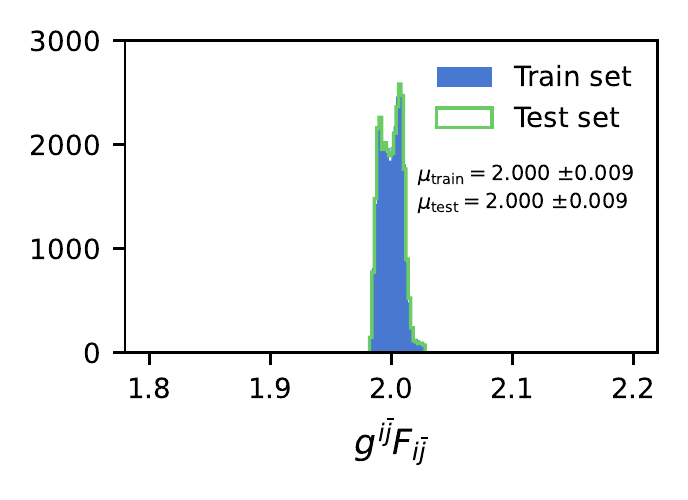}\includegraphics[width=0.3\textwidth]{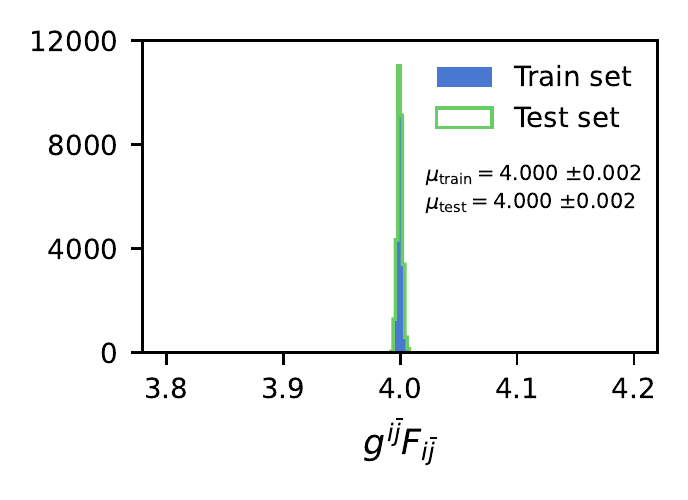}\includegraphics[width=0.3\textwidth]{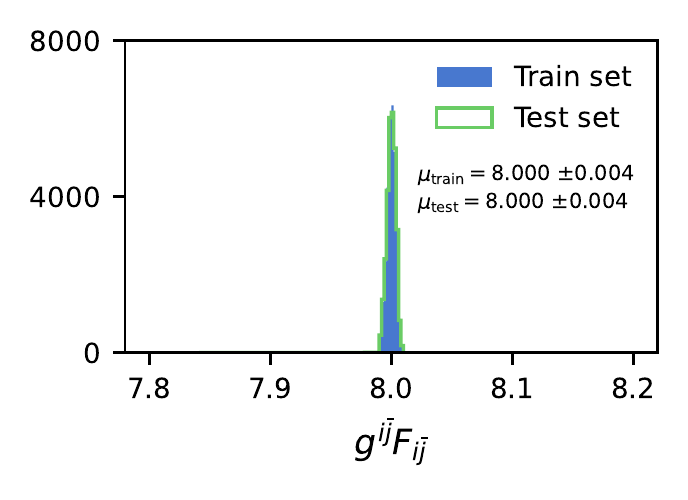}

\caption{Results for line bundle connections on an elliptic curve trained using $\text{Loss}[\boldsymbol{v}]$. The plots show the histogram of $g^{i\bar{j}}F_{i\bar{j}}$ evaluated for sample points on the elliptic curve in both the training and test sets. The left, middle and right plots are for networks of depth $D=2,3,4$ respectively, which correspond to connections on the line bundles $\mathcal{O}_{X}(2)$, $\mathcal{O}_{X}(4)$ and $\mathcal{O}_{X}(8)$.}
\label{fig:elliptic_curve_hists}
\end{figure*}

\subsection{Loss function}

\noindent We now introduce an accuracy measure for HYM connections that will serve as a loss function for the neural network. As a notational convenience, we define the contraction of $g$ with $F$ to be the scalar $F_{g}\equiv g^{i\bar{j}}F_{i\bar{j}}$ valued in endomorphisms of the gauge group; that is,  for a rank-$d$ bundle $V$, at a point on $X$, $F_{g}$ is a $d\times d$ matrix. With this notation the HYM equation is simply 
\begin{equation}
F_{g}=\mu(V)\,\id.
\end{equation}
We also define the expectation $\langle O\rangle$ of a quantity $O$ to be its average over the Calabi--Yau $X$ using the exact CY measure $\vol_{\Omega}$ -- for example, the expectation of $\tr F_{g}$ is defined to be
\begin{equation}
\langle\tr F_{g}\rangle\equiv\int_{X}\vol_{\Omega}\tr F_{g},
\end{equation}
where recall that we normalised $\Vol_{\Omega}\equiv\int_{X}\vol_{\Omega}=1$.

Our connection network outputs $\log G^{-1}$, which in turn is used to compute $F$. Together with the data of an approximate Calabi--Yau metric $g$, this gives $F_{g}$ as a function of the network weights $\boldsymbol{v}$. As we discuss in Appendix \ref{app:loss}, a suitable choice for the loss function of the connection network is
\begin{equation}\label{eq:loss}
\text{Loss}[\boldsymbol{v}]\equiv\langle\tr F_{g}^{2}(\boldsymbol{v})\rangle-\frac{1}{d}\langle\tr F_{g}(\boldsymbol{v})\rangle^{2}.
\end{equation}
Obviously, there are other loss functions that one could choose. For example, given that one can often compute the slope of $V(k)$ by algebraic means, one could instead minimise $\left|\langle\tr F_{g}\rangle-\mu(V(k))\right|$, or any power of this.

\section{Results\label{sec:Results}}

\noindent Having laid out our strategy, we now move to our results. The examples we consider are an elliptic curve, a K3 surface and a quintic threefold, all given as $n$-dimensional hypersurfaces in $\bP^{n+1}$ defined by the zero locus of a degree-$(n+2)$ polynomial of the homogeneous coordinates $[z_0:\ldots:z_{n+1}]$. In all examples, the Ricci-flat metric was first calculated using Douglas et al.'s metric network~\cite{Douglas:2020hpv,MLGeometry}. We then trained the connection network for a variety of network depths. In all examples, the loss function was taken to be $\text{Loss}[\boldsymbol{v}]$ as in (\ref{eq:loss}) and the networks were trained for 500 epochs using the Adam gradient-based optimisation algorithm~\cite{adam}. The training sets and test sets each consisted of 10,000 random sample points on the relevant Calabi--Yau hypersurface. Note that with our choice of normalisations, the slope of the line bundle $V=\mathcal{O}_{X}(m)$ is given by $\mu(V)=m$, and so the HYM connection on $V$ should satisfy $F_{g}=m$.

\newsection{Elliptic curve} The defining equation of an elliptic curve can be expressed in the form\footnote{This can be brought into the usual Weierstrass form $y^2 = x^3 + a x + b$ by defining $x=z_1/z_0$ and $y=z_2/z_0$.}
\begin{equation}\label{eq:ec}
f(z)=z_{1}^{3}+az_{0}^2 z_{1}- z_0 z_2^2+bz_{0}^{3},
\end{equation}
with the curve itself given by the zero locus of $f(z)$ in $\mathbb{P}^2$. The curve is non-singular if and only if the discriminant $\Delta=-16(4a^{3}+27b^{2})$ is non-zero. The example we consider is the elliptic curve with $(a,b)=(-1,1)$. Since $\Delta\neq0$, the curve is smooth and free from singularities. The approximate Calabi--Yau metric for this example was computed at $k=4$ using a network of depth $D=3$ with $W^{(i)}=(70,100,1)$. In the language of \cite{Braun:2007sn}, the accuracy of this metric is $\sigma=0.001$. 

\begin{figure}
\quad\quad\quad\includegraphics[width=0.3\textwidth]{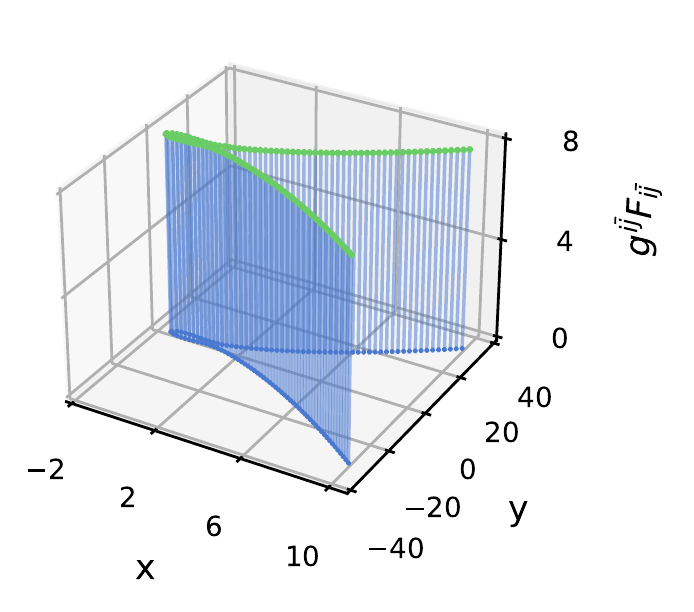}

\caption{Values of $g^{i\bar{j}}F_{i\bar{j}}$ for a line bundle connection calculated using a $D=4$ network on the elliptic curve trained using $\text{Loss}[\boldsymbol{v}]$. The plot shows the values of $g^{i\bar{j}}F_{i\bar{j}}$ on the $z$-axis sampled over points on the elliptic curve on the patch $z_{0}=1$ with $(x,y)=(z_{1},z_{2})$.}
\label{fig:elliptic_curve}
\end{figure}

\begin{figure*}
\includegraphics[width=0.3\textwidth]{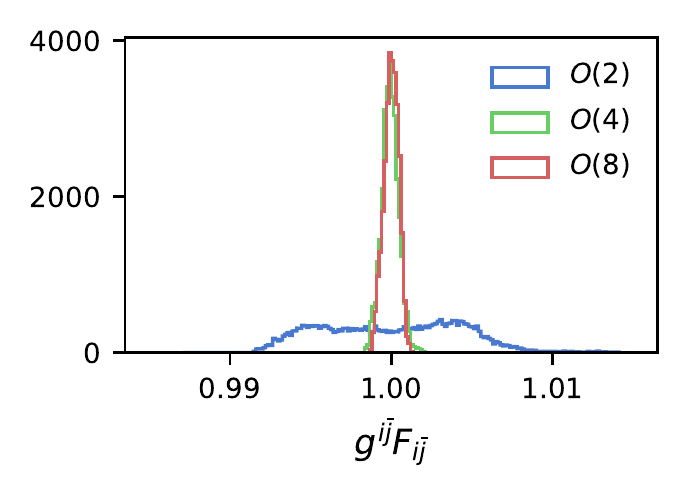}\includegraphics[width=0.3\textwidth]{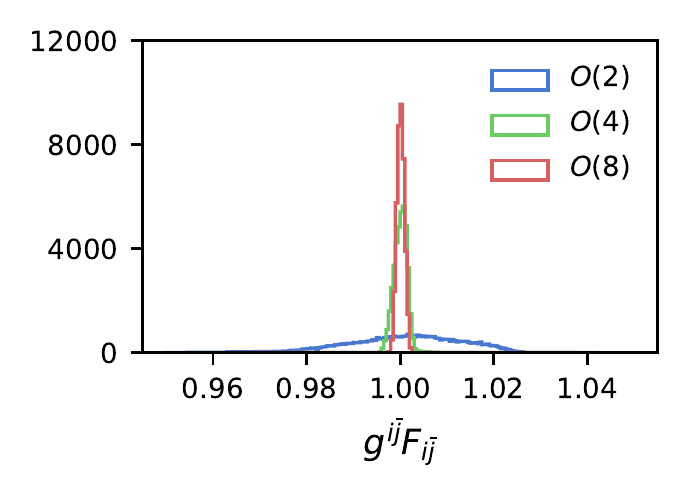}\includegraphics[width=0.295\textwidth]{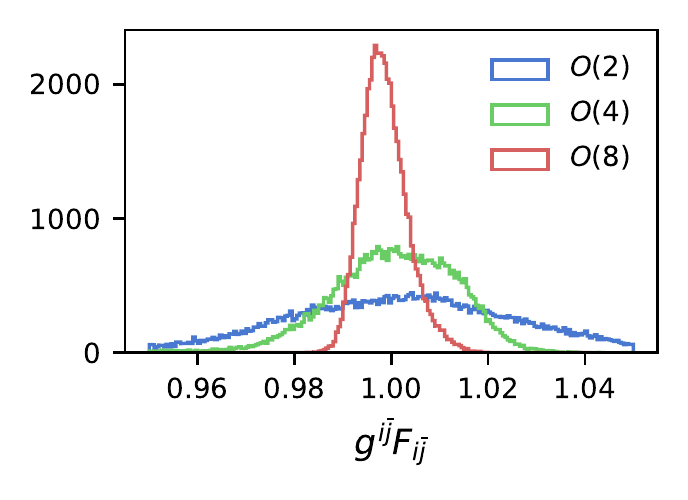}

\caption{Histograms of $g^{i\bar{j}}F_{i\bar{j}}$ on the line bundle $\mathcal{O}_X(1)$ for: (Left) an elliptic curve; (Middle) a K3 surface; (Right) a quintic threefold. In all cases, $g^{i\bar{j}}F_{i\bar{j}}$ is evaluated for points in the test set with the curvature given by untwisting the connection calculated by neural networks with $D=2,3,4$, corresponding to $\mathcal{O}_{X}(2)$, $\mathcal{O}_{X}(4)$ and $\mathcal{O}_{X}(8)$.}
\label{fig:elliptic_K3_quintic_comparison}
\end{figure*}

To compute the connection, we considered neural networks of depth $D=2,3,4$ with intermediate layers of width $W^{(i)}=40$. Our results are shown in \Figref{elliptic_curve_hists} with the training curves given in the appendices in \Figref{ec_training}. In particular, we plot the histogram of $g^{i\bar{j}}F_{i\bar{j}}$ evaluated for points in both the training and test sets. One sees that the histogram of $g^{i\bar{j}}F_{i\bar{j}}$ is clustered around 2, 4 and 8 for the $D=2,3,4$ networks respectively, in agreement with $F_g=m$, with the distribution more peaked for $D=3,4$. In \Figref{elliptic_curve}, we plot the values of $g^{i\bar{j}}F_{i\bar{j}}$ for the $D=4$ network over the elliptic curve on the patch $z_{0}=1$.  As expected from the histograms, the values of $g^{i\bar{j}}F_{i\bar{j}}$ over the elliptic curve are very close to constant.

In order to compare the accuracy of the networks, we use the fact that the curvature of the HYM connection on $\mathcal{O}_{X}(m)$ and that on $\mathcal{O}_{X}(m+k)$ are related in a simple way since they are proportional. As an example, consider $V=\mathcal{O}_{X}(1)$ where we then twist by $\mathcal{O}_{X}(k)$ with $k=1,3,7$. Our neural network then computes the HYM connections on $\mathcal{O}_{X}(2)$, $\mathcal{O}_{X}(4)$ and $\mathcal{O}_{X}(8)$. We then untwist in order to obtain a connection on $V=\mathcal{O}_{X}(1)$ itself. We show the result of this in the left plot of \Figref{elliptic_K3_quintic_comparison}. We see that all three networks are accurate, with the values of $g^{i\bar{j}}F_{i\bar{j}}$ within 1\% of the expected result, i.e.~one, though the $D=2$ network (corresponding to $\mathcal{O}_{X}(2)$) is the least accurate. The deeper $D=3$ and $D=4$ networks, however, have very similar accuracy to each other. This is not that surprising, since the K\"ahler metric on $X$ was itself computed using a $D=3$ network, so the extra freedom allowed by the $D=4$ network is not necessary. 

\newsection{K3 surface} The K3 surface we consider is a smooth quartic hypersurface $f(z) = 0$ in
$\mathbb{P}^3$. The defining equation is
\begin{equation}\label{eq:K3}
f(z)=z_{0}^{4}+z_{1}^{4}+z_{2}^{4}+z_{3}^{4},
\end{equation}
which gives the Fermat quartic. The approximate Calabi--Yau metric for this example was computed at $k=8$ using a network of depth $D=4$ with intermediate layers of width 100, i.e.~$W^{(i)}=(100,100,100,1)$. In the language of \cite{Braun:2007sn}, the resulting metric has a sigma measure of $\sigma=0.00035$.

We considered neural networks of depth $D=2,3,4$ with intermediate layers of width $W^{(i)}=100$. Our full results are given in Appendix \ref{sec:Training-curves}, with the histograms of $g^{i\bar{j}}F_{i\bar{j}}$ evaluated for both the training and test sets shown in \Figref{K3}, and the training curves given in \Figref{K3_training}. One sees that the histogram of $g^{i\bar{j}}F_{i\bar{j}}$ is tightly clustered around 2, 4 and 8 for the $D=2,3,4$ networks respectively, with the distribution more peaked for $D=3,4$. 

In order to compare the accuracy of these three networks, we again treat the networks as computing connections on $\mathcal{O}_X (1+k)$, and then untwist in order to obtain a connection on $V=\mathcal{O}_{X}(1)$. We show the result of this in the middle plot of \Figref{elliptic_K3_quintic_comparison}. We observe that all three networks are accurate, with the values of $g^{i\bar{j}}F_{i\bar{j}}$ within 2\% of the expected result, i.e.~one. Again, the $D=2$ network (corresponding to $\mathcal{O}_{X}(2)$) is the least accurate of the three, as it displays the largest spread in values, and the deepest, $D=4$ network gives the smallest spread of the three. Since the numerical Ricci-flat metric on $X$ was itself computed using a $D=4$ network, the extra complexity allowed by the $D=4$ network does show a small advantage over the $D=3$ network. Note that we computed the approximate Ricci-flat metric using a $D=4$ network -- the curvature of the HYM connection should agree with the K\"ahler form of this metric. It is surprising, therefore, that the $D=4$ connection network does not show an increase in accuracy over $D=3$.

\newsection{Quintic threefold} Finally, we consider a Calabi--Yau threefold given as smooth quintic hypersurface in $\bP^{4}$. The defining equation is
\begin{equation}\label{eq:quintic}
f(z)=z_{0}^{5}+z_{1}^{5}+z_{2}^{5}+z_{3}^{5}+z_{4}^{5}+\tfrac{1}{2}z_{0}z_{1}z_{2}z_{3}z_{4},
\end{equation}
which gives a member of the Dwork family of quintics. The approximate Calabi--Yau metric for this example was computed at $k=8$ using a network of depth $D=4$ with $W^{(i)}=(100,100,100,1)$. The resulting metric has a sigma measure of $\sigma=0.001$.

We considered neural networks of depth $D=2,3,4$ with intermediate layers of width $W^{(i)}=100$. Our full results are given in Appendix \ref{sec:Training-curves}, with the histograms of $g^{i\bar{j}}F_{i\bar{j}}$ evaluated for points in both the training and test sets shown in \Figref{quintic}, and the training curves shown in \Figref{quintic_training}. In order to compare the accuracy of these three networks, we again untwist in order to obtain a connection on $V=\mathcal{O}_{X}(1)$. We show the result of this in the right plot of \Figref{elliptic_K3_quintic_comparison}. We see that all three networks are accurate, with the values of $g^{i\bar{j}}F_{i\bar{j}}$ within 5\% of the expected result, i.e.~one. Note that we computed the approximate Ricci-flat metric using a $D=4$ network, and the curvature of the HYM connection should agree with the K\"ahler form of the Ricci-flat metric. Thanks to this, and the complexity of the Calabi--Yau metric on a threefold, it is not surprising that the $D=4$ network performs the best of the three.

\begin{acknowledgments}
AA is supported by the EU's Horizon 2020 research and innovation program under the Marie Sk\l{}odowska-Curie grant agreement No.~838776. YHH would like to thank STFC for grant ST/J00037X/1. BAO is supported in part by both the research grant DOE No.~DESC0007901 and SAS Account 020-0188-2-010202-6603-0338.
\end{acknowledgments}

\bibliographystyle{utphys}
\bibliography{draft,extra}

\providecommand{\href}[2]{#2}\begingroup\raggedright\begin{thebibliography}{10}

\bibitem{Braun:2005ux}
V.~Braun, Y.-H. He, B.~A. Ovrut, and T.~Pantev, ``{A Heterotic standard
  model}'', \href{http://dx.doi.org/10.1016/j.physletb.2005.05.007}{{\em Phys.
  Lett. B} {\bfseries 618} (2005)252--258},
  \href{http://arxiv.org/abs/hep-th/0501070}{{\ttfamily arXiv:hep-th/0501070}}.

\bibitem{Lukas:1998yy}
A.~Lukas, B.~A. Ovrut, K.~S. Stelle, and D.~Waldram, ``{The Universe as a
  domain wall}'', \href{http://dx.doi.org/10.1103/PhysRevD.59.086001}{{\em
  Phys. Rev. D} {\bfseries 59} (1999)086001},
  \href{http://arxiv.org/abs/hep-th/9803235}{{\ttfamily arXiv:hep-th/9803235}}.

\bibitem{Donagi:1999ez}
R.~Donagi, B.~A. Ovrut, T.~Pantev, and D.~Waldram, ``{Standard models from
  heterotic M theory}'',
  \href{http://dx.doi.org/10.4310/ATMP.2001.v5.n1.a4}{{\em Adv. Theor. Math.
  Phys.} {\bfseries 5} (2002)93--137},
  \href{http://arxiv.org/abs/hep-th/9912208}{{\ttfamily arXiv:hep-th/9912208}}.

\bibitem{Bouchard:2005ag}
V.~Bouchard and R.~Donagi, ``{An SU(5) heterotic standard model}'',
  \href{http://dx.doi.org/10.1016/j.physletb.2005.12.042}{{\em Phys. Lett. B}
  {\bfseries 633} (2006)783--791},
  \href{http://arxiv.org/abs/hep-th/0512149}{{\ttfamily arXiv:hep-th/0512149}}.

\bibitem{Blumenhagen:2006ux}
R.~Blumenhagen, S.~Moster, and T.~Weigand, ``{Heterotic GUT and standard model
  vacua from simply connected Calabi-Yau manifolds}'',
  \href{http://dx.doi.org/10.1016/j.nuclphysb.2006.06.005}{{\em Nucl. Phys. B}
  {\bfseries 751} (2006)186--221},
  \href{http://arxiv.org/abs/hep-th/0603015}{{\ttfamily arXiv:hep-th/0603015}}.

\bibitem{Lebedev:2006kn}
O.~Lebedev, H.~P. Nilles, S.~Raby, S.~Ramos-Sanchez, M.~Ratz, P.~K.~S.
  Vaudrevange, and A.~Wingerter, ``{A Mini-landscape of exact MSSM spectra in
  heterotic orbifolds}'',
  \href{http://dx.doi.org/10.1016/j.physletb.2006.12.012}{{\em Phys. Lett. B}
  {\bfseries 645} (2007)88--94},
  \href{http://arxiv.org/abs/hep-th/0611095}{{\ttfamily arXiv:hep-th/0611095}}.

\bibitem{Candelas:2007ac}
P.~Candelas, X.~de~la Ossa, Y.-H. He, and B.~Szendroi, ``{Triadophilia: A
  Special Corner in the Landscape}'',
  \href{http://dx.doi.org/10.4310/ATMP.2008.v12.n2.a6}{{\em Adv. Theor. Math.
  Phys.} {\bfseries 12} 2, (2008)429--473},
  \href{http://arxiv.org/abs/0706.3134}{{\ttfamily arXiv:0706.3134 [hep-th]}}.

\bibitem{Lebedev:2008un}
O.~Lebedev, H.~P. Nilles, S.~Ramos-Sanchez, M.~Ratz, and P.~K.~S. Vaudrevange,
  ``{Heterotic mini-landscape. (II). Completing the search for MSSM vacua in a
  Z(6) orbifold}'',
  \href{http://dx.doi.org/10.1016/j.physletb.2008.08.054}{{\em Phys. Lett. B}
  {\bfseries 668} (2008)331--335},
  \href{http://arxiv.org/abs/0807.4384}{{\ttfamily arXiv:0807.4384 [hep-th]}}.

\bibitem{MayorgaPena:2012ifg}
D.~K. Mayorga~Pena, H.~P. Nilles, and P.-K. Oehlmann, ``{A Zip-code for Quarks,
  Leptons and Higgs Bosons}'',
  \href{http://dx.doi.org/10.1007/JHEP12(2012)024}{{\em JHEP} {\bfseries 12}
  (2012)024}, \href{http://arxiv.org/abs/1209.6041}{{\ttfamily arXiv:1209.6041
  [hep-th]}}.

\bibitem{Anderson:2009mh}
L.~B. Anderson, J.~Gray, Y.-H. He, and A.~Lukas, ``{Exploring Positive Monad
  Bundles And A New Heterotic Standard Model}'',
  \href{http://dx.doi.org/10.1007/JHEP02(2010)054}{{\em JHEP} {\bfseries 02}
  (2010)054}, \href{http://arxiv.org/abs/0911.1569}{{\ttfamily arXiv:0911.1569
  [hep-th]}}.

\bibitem{Anderson:2011ns}
L.~B. Anderson, J.~Gray, A.~Lukas, and E.~Palti, ``{Two Hundred Heterotic
  Standard Models on Smooth Calabi-Yau Threefolds}'',
  \href{http://dx.doi.org/10.1103/PhysRevD.84.106005}{{\em Phys. Rev. D}
  {\bfseries 84} (2011)106005},
  \href{http://arxiv.org/abs/1106.4804}{{\ttfamily arXiv:1106.4804 [hep-th]}}.

\bibitem{Anderson:2013xka}
L.~B. Anderson, A.~Constantin, J.~Gray, A.~Lukas, and E.~Palti, ``{A
  Comprehensive Scan for Heterotic SU(5) GUT models}'',
  \href{http://dx.doi.org/10.1007/JHEP01(2014)047}{{\em JHEP} {\bfseries 01}
  (2014)047}, \href{http://arxiv.org/abs/1307.4787}{{\ttfamily arXiv:1307.4787
  [hep-th]}}.

\bibitem{Donaldson}
S.~K. Donaldson, ``Anti self-dual yang-mills connections over complex algebraic
  surfaces and stable vector bundles'',
  \href{http://dx.doi.org/10.1112/plms/s3-50.1.1}{{\em Proceedings of the
  London Mathematical Society} {\bfseries s3-50} 1, (1985)1--26}.

\bibitem{UhlenbeckYau}
K.~Uhlenbeck and S.~T. Yau, ``On the existence of hermitian-yang-mills
  connections in stable vector bundles'',
  \href{http://dx.doi.org/10.1002/cpa.3160390714}{{\em Communications on Pure
  and Applied Mathematics} {\bfseries 39} S1, (1986)S257--S293}.

\bibitem{Greene:1986ar}
B.~R. Greene, K.~H. Kirklin, P.~J. Miron, and G.~G. Ross, ``{A Superstring
  Inspired Standard Model}'',
  \href{http://dx.doi.org/10.1016/0370-2693(86)90137-1}{{\em Phys. Lett. B}
  {\bfseries 180} (1986)69}.

\bibitem{Greene:1986bm}
B.~R. Greene, K.~H. Kirklin, P.~J. Miron, and G.~G. Ross, ``{A Three Generation
  Superstring Model. 1. Compactification and Discrete Symmetries}'',
  \href{http://dx.doi.org/10.1016/0550-3213(86)90057-X}{{\em Nucl. Phys. B}
  {\bfseries 278} (1986)667--693}.

\bibitem{Greene:1986jb}
B.~R. Greene, K.~H. Kirklin, P.~J. Miron, and G.~G. Ross, ``{A Three Generation
  Superstring Model. 2. Symmetry Breaking and the Low-Energy Theory}'',
  \href{http://dx.doi.org/10.1016/0550-3213(87)90662-6}{{\em Nucl. Phys. B}
  {\bfseries 292} (1987)606--652}.

\bibitem{Matsuoka:1986vg}
T.~Matsuoka and D.~Suematsu, ``{Realistic Models From the $E(8)$ X $E(8)$-prime
  Superstring Theory}'', \href{http://dx.doi.org/10.1143/PTP.76.886}{{\em Prog.
  Theor. Phys.} {\bfseries 76} (1986)886}.

\bibitem{Greene:1987xh}
B.~R. Greene, K.~H. Kirklin, P.~J. Miron, and G.~G. Ross, ``{27**3 Yukawa
  Couplings for a Three Generation Superstring Model}'',
  \href{http://dx.doi.org/10.1016/0370-2693(87)91151-8}{{\em Phys. Lett. B}
  {\bfseries 192} (1987)111--118}.

\bibitem{Donagi:2000zs}
R.~Donagi, B.~A. Ovrut, T.~Pantev, and D.~Waldram, ``{Standard model
  bundles}'', \href{http://dx.doi.org/10.4310/ATMP.2001.v5.n3.a5}{{\em Adv.
  Theor. Math. Phys.} {\bfseries 5} (2002)563--615},
  \href{http://arxiv.org/abs/math/0008010}{{\ttfamily arXiv:math/0008010}}.

\bibitem{Braun:2006me}
V.~Braun, Y.-H. He, and B.~A. Ovrut, ``{Yukawa couplings in heterotic standard
  models}'', \href{http://dx.doi.org/10.1088/1126-6708/2006/04/019}{{\em JHEP}
  {\bfseries 04} (2006)019},
  \href{http://arxiv.org/abs/hep-th/0601204}{{\ttfamily arXiv:hep-th/0601204}}.

\bibitem{Anderson:2010tc}
L.~B. Anderson, J.~Gray, and B.~Ovrut, ``{Yukawa Textures From Heterotic
  Stability Walls}'', \href{http://dx.doi.org/10.1007/JHEP05(2010)086}{{\em
  JHEP} {\bfseries 05} (2010)086},
  \href{http://arxiv.org/abs/1001.2317}{{\ttfamily arXiv:1001.2317 [hep-th]}}.

\bibitem{Headrick:2005ch}
M.~Headrick and T.~Wiseman, ``{Numerical Ricci-flat metrics on K3}'',
  \href{http://dx.doi.org/10.1088/0264-9381/22/23/002}{{\em Class. Quant.
  Grav.} {\bfseries 22} (2005)4931--4960},
  \href{http://arxiv.org/abs/hep-th/0506129}{{\ttfamily arXiv:hep-th/0506129}}.

\bibitem{Douglas:2006rr}
M.~R. Douglas, R.~L. Karp, S.~Lukic, and R.~Reinbacher, ``{Numerical Calabi-Yau
  metrics}'', \href{http://dx.doi.org/10.1063/1.2888403}{{\em J. Math. Phys.}
  {\bfseries 49} (2008)032302},
  \href{http://arxiv.org/abs/hep-th/0612075}{{\ttfamily arXiv:hep-th/0612075}}.

\bibitem{Braun:2007sn}
V.~Braun, T.~Brelidze, M.~R. Douglas, and B.~A. Ovrut, ``{Calabi-Yau Metrics
  for Quotients and Complete Intersections}'',
  \href{http://dx.doi.org/10.1088/1126-6708/2008/05/080}{{\em JHEP} {\bfseries
  05} (2008)080}, \href{http://arxiv.org/abs/0712.3563}{{\ttfamily
  arXiv:0712.3563 [hep-th]}}.

\bibitem{Headrick:2009jz}
M.~Headrick and A.~Nassar, ``{Energy functionals for Calabi-Yau metrics}'',
  \href{http://dx.doi.org/10.4310/ATMP.2013.v17.n5.a1}{{\em Adv. Theor. Math.
  Phys.} {\bfseries 17} 5, (2013)867--902},
  \href{http://arxiv.org/abs/0908.2635}{{\ttfamily arXiv:0908.2635 [hep-th]}}.

\bibitem{Tian}
G.~Tian, ``{On a set of polarized Kähler metrics on algebraic manifolds}'',
  \href{http://dx.doi.org/10.4310/jdg/1214445039}{{\em Journal of Differential
  Geometry} {\bfseries 32} 1, (1990)99 -- 130}.

\bibitem{math/0512625}
S.~K. {Donaldson}, ``{Some numerical results in complex differential
  geometry}'', \href{http://arxiv.org/abs/math/0512625}{{\ttfamily
  arXiv:math/0512625 [math.DG]}}.

\bibitem{Ashmore:2019wzb}
A.~Ashmore, Y.-H. He, and B.~A. Ovrut, ``{Machine Learning
  Calabi\textendash{}Yau Metrics}'',
  \href{http://dx.doi.org/10.1002/prop.202000068}{{\em Fortsch. Phys.}
  {\bfseries 68} 9, (2020)2000068},
  \href{http://arxiv.org/abs/1910.08605}{{\ttfamily arXiv:1910.08605
  [hep-th]}}.

\bibitem{Anderson:2020hux}
L.~B. Anderson, M.~Gerdes, J.~Gray, S.~Krippendorf, N.~Raghuram, and F.~Ruehle,
  ``{Moduli-dependent Calabi-Yau and SU(3)-structure metrics from Machine
  Learning}'', \href{http://dx.doi.org/10.1007/JHEP05(2021)013}{{\em JHEP}
  {\bfseries 05} (2021)013}, \href{http://arxiv.org/abs/2012.04656}{{\ttfamily
  arXiv:2012.04656 [hep-th]}}.

\bibitem{Douglas:2020hpv}
M.~R. Douglas, S.~Lakshminarasimhan, and Y.~Qi, ``{Numerical Calabi-Yau metrics
  from holomorphic networks}'',
  \href{http://arxiv.org/abs/2012.04797}{{\ttfamily arXiv:2012.04797
  [hep-th]}}.

\bibitem{Jejjala:2020wcc}
V.~Jejjala, D.~K. Mayorga~Pena, and C.~Mishra, ``{Neural Network Approximations
  for Calabi-Yau Metrics}'', \href{http://arxiv.org/abs/2012.15821}{{\ttfamily
  arXiv:2012.15821 [hep-th]}}.

\bibitem{Douglas:2021zdn}
M.~R. Douglas, ``{Holomorphic feedforward networks}'',
  \href{http://arxiv.org/abs/2105.03991}{{\ttfamily arXiv:2105.03991
  [math.CV]}}.

\bibitem{He:2018jtw}
Y.-H. He, \href{http://dx.doi.org/10.1007/978-3-030-77562-9}{{\em {The
  Calabi\textendash{}Yau Landscape: From Geometry, to Physics, to Machine
  Learning}}}.
\newblock Lecture Notes in Mathematics. 5, 2021.
\newblock \href{http://arxiv.org/abs/1812.02893}{{\ttfamily arXiv:1812.02893
  [hep-th]}}.

\bibitem{Douglas:2006hz}
M.~R. Douglas, R.~L. Karp, S.~Lukic, and R.~Reinbacher, ``{Numerical solution
  to the hermitian Yang-Mills equation on the Fermat quintic}'',
  \href{http://dx.doi.org/10.1088/1126-6708/2007/12/083}{{\em JHEP} {\bfseries
  12} (2007)083}, \href{http://arxiv.org/abs/hep-th/0606261}{{\ttfamily
  arXiv:hep-th/0606261}}.

\bibitem{Anderson:2010ke}
L.~B. Anderson, V.~Braun, R.~L. Karp, and B.~A. Ovrut, ``{Numerical Hermitian
  Yang-Mills Connections and Vector Bundle Stability in Heterotic Theories}'',
  \href{http://dx.doi.org/10.1007/JHEP06(2010)107}{{\em JHEP} {\bfseries 06}
  (2010)107}, \href{http://arxiv.org/abs/1004.4399}{{\ttfamily arXiv:1004.4399
  [hep-th]}}.

\bibitem{Anderson:2011ed}
L.~B. Anderson, V.~Braun, and B.~A. Ovrut, ``{Numerical Hermitian Yang-Mills
  Connections and Kahler Cone Substructure}'',
  \href{http://dx.doi.org/10.1007/JHEP01(2012)014}{{\em JHEP} {\bfseries 01}
  (2012)014}, \href{http://arxiv.org/abs/1103.3041}{{\ttfamily arXiv:1103.3041
  [hep-th]}}.

\bibitem{Wang}
X.~Wang, ``{Canonical metrics on stable vector bundles}'',
  \href{http://dx.doi.org/10.4310/CAG.2005.v13.n2.a1}{{\em Communications in
  Analysis and Geometry} {\bfseries 13} 2, (2005)253 -- 285}.

\bibitem{Anderson:2012yf}
L.~B. Anderson, J.~Gray, A.~Lukas, and E.~Palti, ``{Heterotic Line Bundle
  Standard Models}'', \href{http://dx.doi.org/10.1007/JHEP06(2012)113}{{\em
  JHEP} {\bfseries 06} (2012)113},
  \href{http://arxiv.org/abs/1202.1757}{{\ttfamily arXiv:1202.1757 [hep-th]}}.

\bibitem{GrootNibbelink:2015lme}
S.~Groot~Nibbelink, O.~Loukas, and F.~Ruehle, ``{(MS)SM-like models on smooth
  Calabi-Yau manifolds from all three heterotic string theories}'',
  \href{http://dx.doi.org/10.1002/prop.201500041}{{\em Fortsch. Phys.}
  {\bfseries 63} (2015)609--632},
  \href{http://arxiv.org/abs/1507.07559}{{\ttfamily arXiv:1507.07559
  [hep-th]}}.

\bibitem{GrootNibbelink:2015dvi}
S.~Groot~Nibbelink, O.~Loukas, F.~Ruehle, and P.~K.~S. Vaudrevange, ``{Infinite
  number of MSSMs from heterotic line bundles?}'',
  \href{http://dx.doi.org/10.1103/PhysRevD.92.046002}{{\em Phys. Rev. D}
  {\bfseries 92} 4, (2015)046002},
  \href{http://arxiv.org/abs/1506.00879}{{\ttfamily arXiv:1506.00879
  [hep-th]}}.

\bibitem{GrootNibbelin:2016ovb}
S.~Groot~Nibbelin and F.~Ruehle, ``{Line bundle embeddings for heterotic
  theories}'', \href{http://dx.doi.org/10.1007/JHEP04(2016)186}{{\em JHEP}
  {\bfseries 04} (2016)186}, \href{http://arxiv.org/abs/1601.00676}{{\ttfamily
  arXiv:1601.00676 [hep-th]}}.

\bibitem{Braun:2017feb}
A.~P. Braun, C.~R. Brodie, and A.~Lukas, ``{Heterotic Line Bundle Models on
  Elliptically Fibered Calabi-Yau Three-folds}'',
  \href{http://dx.doi.org/10.1007/JHEP04(2018)087}{{\em JHEP} {\bfseries 04}
  (2018)087}, \href{http://arxiv.org/abs/1706.07688}{{\ttfamily
  arXiv:1706.07688 [hep-th]}}.

\bibitem{MLGeometry}
M.~R. Douglas, S.~Lakshminarasimhan, and Y.~Qi, {\em MLGeometry}, 2021.
\newblock \url{https://github.com/yidiq7/MLGeometry}.

\bibitem{hep-th/0512177}
V.~Braun, Y.-H. He, B.~A. Ovrut, and T.~Pantev, ``{The Exact MSSM spectrum from
  string theory}'', \href{http://dx.doi.org/10.1088/1126-6708/2006/05/043}{{\em
  JHEP} {\bfseries 05} (2006)043},
  \href{http://arxiv.org/abs/hep-th/0512177}{{\ttfamily arXiv:hep-th/0512177}}.

\bibitem{hep-th/0502155}
V.~Braun, Y.-H. He, B.~A. Ovrut, and T.~Pantev, ``{A Standard model from the
  E(8) x E(8) heterotic superstring}'',
  \href{http://dx.doi.org/10.1088/1126-6708/2005/06/039}{{\em JHEP} {\bfseries
  06} (2005)039}, \href{http://arxiv.org/abs/hep-th/0502155}{{\ttfamily
  arXiv:hep-th/0502155}}.

\bibitem{hep-th/0501070}
V.~Braun, Y.-H. He, B.~A. Ovrut, and T.~Pantev, ``{A Heterotic standard
  model}'', \href{http://dx.doi.org/10.1016/j.physletb.2005.05.007}{{\em Phys.
  Lett. B} {\bfseries 618} (2005)252--258},
  \href{http://arxiv.org/abs/hep-th/0501070}{{\ttfamily arXiv:hep-th/0501070}}.

\bibitem{hep-th/0512149}
V.~Bouchard and R.~Donagi, ``{An SU(5) heterotic standard model}'',
  \href{http://dx.doi.org/10.1016/j.physletb.2005.12.042}{{\em Phys. Lett. B}
  {\bfseries 633} (2006)783--791},
  \href{http://arxiv.org/abs/hep-th/0512149}{{\ttfamily arXiv:hep-th/0512149}}.

\bibitem{0911.1569}
L.~B. Anderson, J.~Gray, Y.-H. He, and A.~Lukas, ``{Exploring Positive Monad
  Bundles And A New Heterotic Standard Model}'',
  \href{http://dx.doi.org/10.1007/JHEP02(2010)054}{{\em JHEP} {\bfseries 02}
  (2010)054}, \href{http://arxiv.org/abs/0911.1569}{{\ttfamily arXiv:0911.1569
  [hep-th]}}.

\bibitem{1112.1097}
V.~Braun, P.~Candelas, R.~Davies, and R.~Donagi, ``{The MSSM Spectrum from
  (0,2)-Deformations of the Heterotic Standard Embedding}'',
  \href{http://dx.doi.org/10.1007/JHEP05(2012)127}{{\em JHEP} {\bfseries 05}
  (2012)127}, \href{http://arxiv.org/abs/1112.1097}{{\ttfamily arXiv:1112.1097
  [hep-th]}}.

\bibitem{1106.4804}
L.~B. Anderson, J.~Gray, A.~Lukas, and E.~Palti, ``{Two Hundred Heterotic
  Standard Models on Smooth Calabi-Yau Threefolds}'',
  \href{http://dx.doi.org/10.1103/PhysRevD.84.106005}{{\em Phys. Rev. D}
  {\bfseries 84} (2011)106005},
  \href{http://arxiv.org/abs/1106.4804}{{\ttfamily arXiv:1106.4804 [hep-th]}}.

\bibitem{1202.1757}
L.~B. Anderson, J.~Gray, A.~Lukas, and E.~Palti, ``{Heterotic Line Bundle
  Standard Models}'', \href{http://dx.doi.org/10.1007/JHEP06(2012)113}{{\em
  JHEP} {\bfseries 06} (2012)113},
  \href{http://arxiv.org/abs/1202.1757}{{\ttfamily arXiv:1202.1757 [hep-th]}}.

\bibitem{1307.4787}
L.~B. Anderson, A.~Constantin, J.~Gray, A.~Lukas, and E.~Palti, ``{A
  Comprehensive Scan for Heterotic SU(5) GUT models}'',
  \href{http://dx.doi.org/10.1007/JHEP01(2014)047}{{\em JHEP} {\bfseries 01}
  (2014)047}, \href{http://arxiv.org/abs/1307.4787}{{\ttfamily arXiv:1307.4787
  [hep-th]}}.

\bibitem{1506.00879}
S.~Groot~Nibbelink, O.~Loukas, F.~Ruehle, and P.~K.~S. Vaudrevange, ``{Infinite
  number of MSSMs from heterotic line bundles?}'',
  \href{http://dx.doi.org/10.1103/PhysRevD.92.046002}{{\em Phys. Rev. D}
  {\bfseries 92} 4, (2015)046002},
  \href{http://arxiv.org/abs/1506.00879}{{\ttfamily arXiv:1506.00879
  [hep-th]}}.

\bibitem{1507.07559}
S.~Groot~Nibbelink, O.~Loukas, and F.~Ruehle, ``{(MS)SM-like models on smooth
  Calabi-Yau manifolds from all three heterotic string theories}'',
  \href{http://dx.doi.org/10.1002/prop.201500041}{{\em Fortsch. Phys.}
  {\bfseries 63} (2015)609--632},
  \href{http://arxiv.org/abs/1507.07559}{{\ttfamily arXiv:1507.07559
  [hep-th]}}.

\bibitem{1007.0203}
M.~Blaszczyk, S.~Groot~Nibbelink, F.~Ruehle, M.~Trapletti, and P.~K.~S.
  Vaudrevange, ``{Heterotic MSSM on a Resolved Orbifold}'',
  \href{http://dx.doi.org/10.1007/JHEP09(2010)065}{{\em JHEP} {\bfseries 09}
  (2010)065}, \href{http://arxiv.org/abs/1007.0203}{{\ttfamily arXiv:1007.0203
  [hep-th]}}.

\bibitem{hep-th/9903052}
B.~Andreas, G.~Curio, and A.~Klemm, ``{Towards the Standard Model spectrum from
  elliptic Calabi-Yau}'',
  \href{http://dx.doi.org/10.1142/S0217751X04018087}{{\em Int. J. Mod. Phys. A}
  {\bfseries 19} (2004)1987},
  \href{http://arxiv.org/abs/hep-th/9903052}{{\ttfamily arXiv:hep-th/9903052}}.

\bibitem{Braun:2008jp}
V.~Braun, T.~Brelidze, M.~R. Douglas, and B.~A. Ovrut, ``{Eigenvalues and
  Eigenfunctions of the Scalar Laplace Operator on Calabi-Yau Manifolds}'',
  \href{http://dx.doi.org/10.1088/1126-6708/2008/07/120}{{\em JHEP} {\bfseries
  07} (2008)120}, \href{http://arxiv.org/abs/0805.3689}{{\ttfamily
  arXiv:0805.3689 [hep-th]}}.

\bibitem{Ashmore:2020ujw}
A.~Ashmore, ``{Eigenvalues and eigenforms on Calabi-Yau threefolds}'',
  \href{http://arxiv.org/abs/2011.13929}{{\ttfamily arXiv:2011.13929
  [hep-th]}}.

\bibitem{Afkhami-Jeddi:2021qkf}
N.~Afkhami-Jeddi, A.~Ashmore, and C.~Cordova, ``{Calabi-Yau CFTs and Random
  Matrices}'', \href{http://arxiv.org/abs/2107.11461}{{\ttfamily
  arXiv:2107.11461 [hep-th]}}.

\bibitem{hep-th/9902071}
A.~Lukas, B.~A. Ovrut, and D.~Waldram, ``{Boundary inflation}'',
  \href{http://dx.doi.org/10.1103/PhysRevD.61.023506}{{\em Phys. Rev. D}
  {\bfseries 61} (2000)023506},
  \href{http://arxiv.org/abs/hep-th/9902071}{{\ttfamily arXiv:hep-th/9902071}}.

\bibitem{McOrist:2016cfl}
J.~McOrist, ``{On the Effective Field Theory of Heterotic Vacua}'',
  \href{http://dx.doi.org/10.1007/s11005-017-1025-0}{{\em Lett. Math. Phys.}
  {\bfseries 108} 4, (2018)1031--1081},
  \href{http://arxiv.org/abs/1606.05221}{{\ttfamily arXiv:1606.05221
  [hep-th]}}.

\bibitem{1801.09645}
c.~Blesneag, E.~I. Buchbinder, A.~Constantin, A.~Lukas, and E.~Palti, ``{Matter
  field K\"ahler metric in heterotic string theory from localisation}'',
  \href{http://dx.doi.org/10.1007/JHEP04(2018)139}{{\em JHEP} {\bfseries 04}
  (2018)139}, \href{http://arxiv.org/abs/1801.09645}{{\ttfamily
  arXiv:1801.09645 [hep-th]}}.

\bibitem{Ishiguro:2021drk}
K.~Ishiguro, T.~Kobayashi, and H.~Otsuka, ``{Hierarchical structure of physical
  Yukawa couplings from matter field K\"ahler metric}'',
  \href{http://dx.doi.org/10.1007/JHEP07(2021)064}{{\em JHEP} {\bfseries 07}
  (2021)064}, \href{http://arxiv.org/abs/2103.10240}{{\ttfamily
  arXiv:2103.10240 [hep-th]}}.

\bibitem{adam}
D.~P. {Kingma} and J.~{Ba}, ``{Adam: A Method for Stochastic Optimization}'',
  \href{http://arxiv.org/abs/1412.6980}{{\ttfamily arXiv:1412.6980 [cs.LG]}}.

\end{thebibliography}\endgroup

\clearpage

\appendix

\section{The network structure and loss function}\label{app:loss}

In this appendix, we further discuss the structure of the neural network for computing numerical connections on line bundles and also justify the choice of loss function in the main text.

\subsection{The network structure}

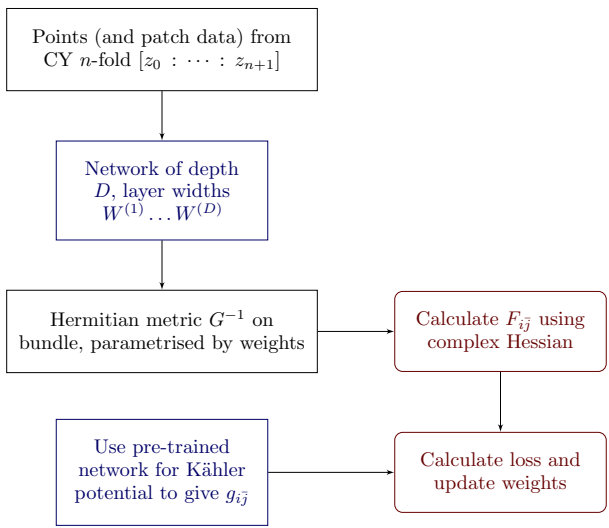
\begin{figure}
  \scalebox{0.75}{
\begin{tikzpicture}[node distance = 2.5cm, auto] 
    \node [block] (b1) {Points (and patch data) from CY $n$-fold $[z_0 : \dots : z_{n+1} ]$};  
    \node [netblock, below of = b1](b3){Network of depth $D$, layer widths $W^{(1)} \dots W^{(D)}$};  
    \node [block, below of = b3](b4){Hermitian metric $G^{-1}$ on bundle, parametrised by weights};
    \node [netblock2, right of =b4, xshift=3.5cm](b5){Calculate $F_{i\bar{j}}$ using complex Hessian};
    \node [netblock, below of = b4, xshift=0cm](b6){Use pre-trained network for K\"ahler potential to give $g_{i\bar{j}}$};
    \node [netblock2, below of = b5](b7){Calculate loss and update weights};
    \path [line] (b1) -- (b3);   
    \path [line] (b3) -- (b4);
    \path [line] (b4) -- (b5);
    \path [line] (b5) -- (b7);
    \path [line] (b6) -- (b7);
  \end{tikzpicture}
}
\caption{Network structure for determining the gauge connection of a line bundle on a CY $n$-fold.}
\label{fig:connection_network}
\end{figure}

The network structure for determining a gauge connection that satisfies hermitian Yang--Mills is given in Figure \ref{fig:connection_network}. As we also describe in the main text, the inputs to the network are sets of points on the Calabi--Yau hypersurface, given as points on the ambient projective space. This information is fed into a linear, dense network of depth $D$ with layer widths $W^{(i)}$, shown in Figure \ref{fig:linear_network}. The first layer of the network is a bihomogeneous layer, followed by dense layers with quadratic activation functions and zero biases. The output of the network is the hermitian metric on the bundle (actually $\log G^{-1}$), parametrised by the weights $\boldsymbol v$ of the network. Next, we compute the curvature $F_{i\bar j}$ of the connection induced by this hermitian metric by taking the complex Hessian of the network. In practice, this computes the curvature as a tensor on the ambient space, so one must pull it back to the hypersurface using the Jacobian defined by the patches and the defining equation of the Calabi--Yau (see \cite{Douglas:2006rr} for more details about this). The loss function of the network, whose discretised form is given in Equation (\ref{eq:loss_discrete}), is defined by both the curvature computed by the network and a numerical Calabi--Yau metric. The latter comes from a pre-trained network whose output is a K\"ahler metric $g_{i\bar j}$ which is approximately Ricci flat. The loss function is then minimised by adjusting the weights $\boldsymbol v$ of the network using backpropagation. The result of this is a ``trained'' network whose output defines an approximate hermitian Yang--Mills connection.

\subsection{The loss function}

Consider the variance of the trace of $F_{g}$:
\begin{equation}
\text{Var}[\tr F_{g}]  =\langle(\tr F_{g})^{2}\rangle-\langle\tr F_{g}\rangle^{2}.
\end{equation}
Clearly, $\text{Var}[\tr F_{g}]\geq0$ with equality only when $\tr F_{g}=\langle\tr F_{g}\rangle$. However, this is the case only if $\tr F_{g}$ is \emph{constant} over $X$. Now consider the fact that for a $d\times d$ hermitian matrix $M$ (such as $\langle\tr F_{g}\rangle$), one always has
\begin{equation}
d\cdot\tr M^{2}\geq(\tr M)^{2},\label{eq:hermitian_inequality}
\end{equation}
with equality if and only if $M$ is proportional to the identity matrix, $M\propto\id$ (but with no constraint on the function relating the two). Putting together these two observations, we define the functional
\begin{equation}
E[F,g]=\langle\tr F_{g}^{2}\rangle-\frac{1}{d}\langle\tr F_{g}\rangle^{2},
\end{equation}
which satisfies 
\begin{equation}
    0\leq\text{Var}[\tr F_{g}]\leq d\cdot E[F,g].
\end{equation}
Thus if one finds a connection such that $E[F,g]=0$, it must be the case that (\ref{eq:hermitian_inequality}) is saturated and $\text{Var}[\tr F_{g}]=0$, which imply $\tr F_{g}=\langle\tr F_{g}\rangle$ and $F_{g}\propto\id$ respectively. Taken together, these two conditions are equivalent to $F_{g}=c\,\id$ with $c$ constant, and so one has found a HYM connection. Conversely, it is clear that a HYM connection satisfies both $\op{Var}[\tr F_{g}]=0$ and saturates the inequality (\ref{eq:hermitian_inequality}). Hence, we have
\begin{equation}
\text{\ensuremath{F} solves HYM}\quad\Leftrightarrow\quad E[F,g]=0.
\end{equation}
This motivates our choice of loss function, $\text{Loss}[\boldsymbol{v}]$, in Equation (\ref{eq:loss}). For the special case of a rank-one bundle, i.e.~a line bundle with $d=1$, one has 
\begin{equation}
E[F,g]=\langle F_{g}^{2}\rangle-\langle F_{g}\rangle^{2},
\end{equation}
which is simply the variance of $F_{g}$.

\begin{figure*}
\includegraphics[width=0.3\textwidth]{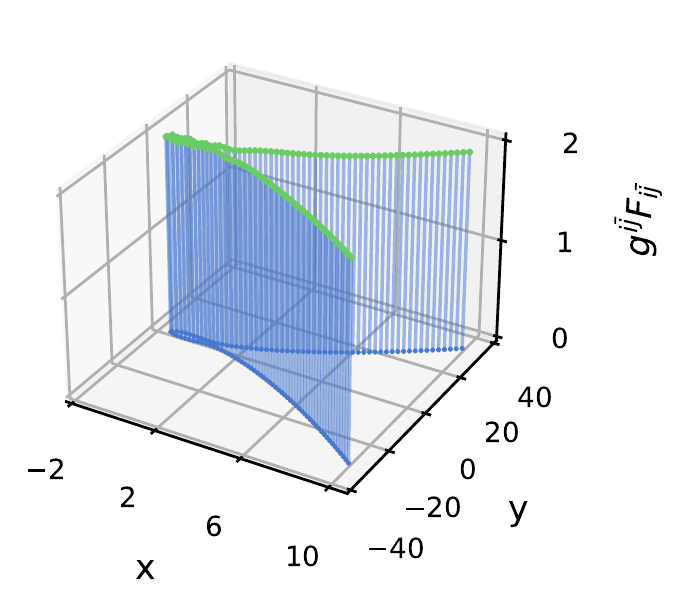}\includegraphics[width=0.3\textwidth]{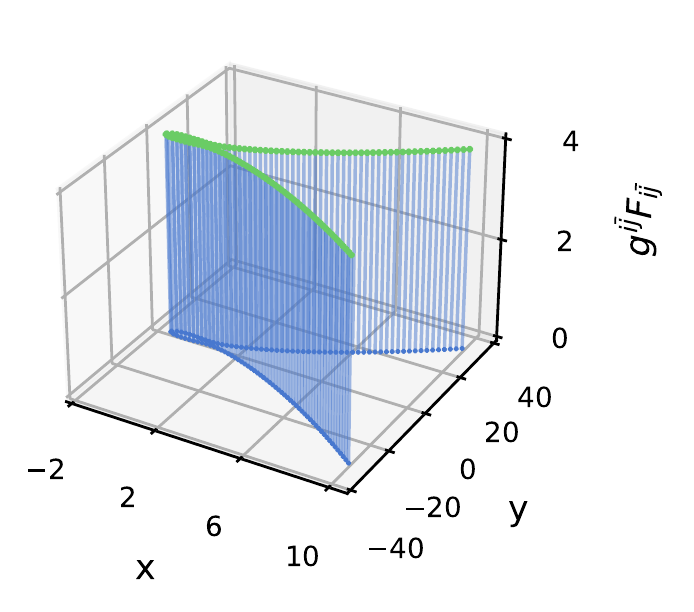}\includegraphics[width=0.3\textwidth]{images/ec3_hm_O8_s1_curve}

\caption{Results for line bundle connections on an elliptic curve trained using $\text{Loss}[\boldsymbol{v}]$. The plots show the values of $g^{i\bar{j}}F_{i\bar{j}}$ on the $z$-axis over the elliptic curve on the patch $z_{0}=1$ with $(x,y)=(z_{1},z_{2})$. The left, middle and right plots are for networks of depth $D=2,3,4$ respectively, which correspond to connections on the line bundles $\mathcal{O}_{X}(2)$, $\mathcal{O}_{X}(4)$ and $\mathcal{O}_{X}(8)$.}
\label{fig:all_elliptic_curve}
\end{figure*}

\begin{figure*}
\includegraphics[width=0.3\textwidth]{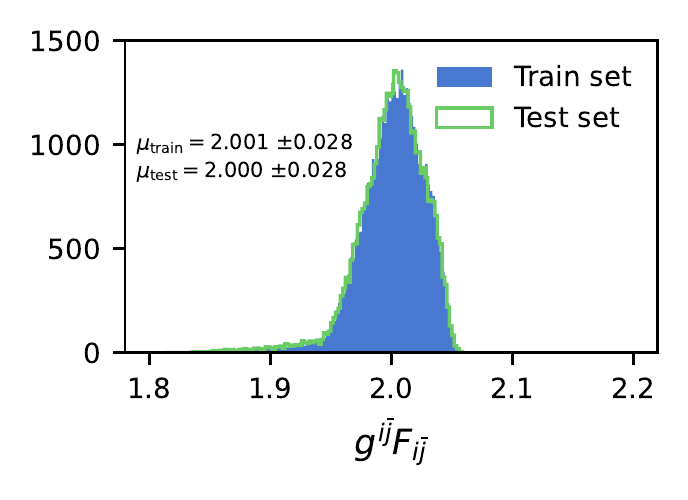}\includegraphics[width=0.3\textwidth]{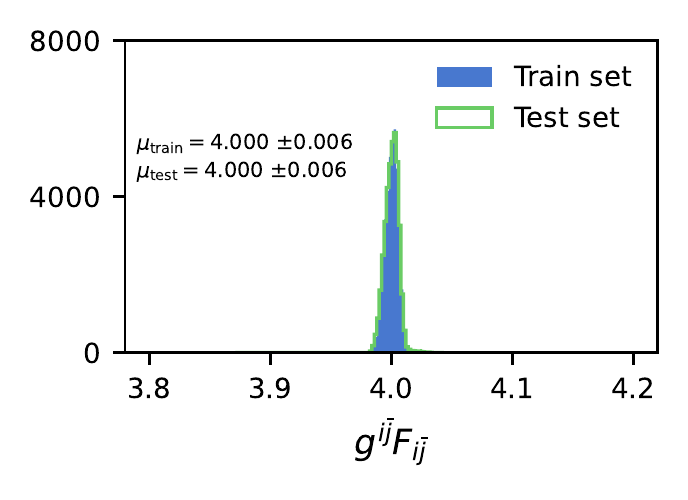}\includegraphics[width=0.3\textwidth]{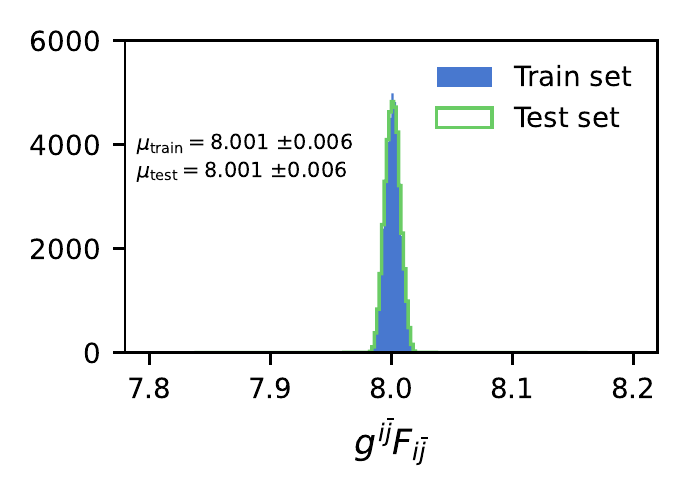}

\caption{Results for line bundle connections on a K3 surface trained using $\text{Loss}[\boldsymbol{v}]$. The plots show the histogram of $g^{i\bar{j}}F_{i\bar{j}}$ evaluated for points in both the training and test sets, together with the mean and standard deviation. The plots are for networks of depth $D=2,3,4$ which correspond to connections on the line bundles $\mathcal{O}_{X}(2)$, $\mathcal{O}_{X}(4)$ and $\mathcal{O}_{X}(8)$.}
\label{fig:K3}
\end{figure*}

Written as a discrete sum over points of $X$, the loss function is
\begin{equation}\label{eq:loss_discrete}
\begin{split}
\text{Loss}[\boldsymbol{v}]&=\frac{\sum_{p}\tr\left(g^{i\bar{j}}(p)F_{i\bar{j}}(\boldsymbol{v},p)\right)^{2}w_{p}}{\sum_{p}w_{p}}\\
&\eqspace -\frac{1}{d}\frac{\left(\sum_{p}g^{i\bar{j}}(p)\tr F_{i\bar{j}}(\boldsymbol{v},p)w_{p}\right)^{2}}{\left(\sum_{p}w_{p}\right)^{2}},
\end{split}
\end{equation}
where $p$ denotes a point in the training set, and $w_{p}$ is a mass which weights the sum over points to reproduce the integration measure defined by $\vol_{\Omega}$~\cite{Douglas:2006rr}.

\section{Histograms and training curves\label{sec:Training-curves}}

In this appendix, we display the histograms for $g^{i\bar j} F_{i \bar j}$ and the training curves for $D=2,3,4$ networks computing line bundle connections on an elliptic curve, K3 surface, and quintic threefold, as in the main text.

\begin{figure*}
\includegraphics[width=0.3\textwidth]{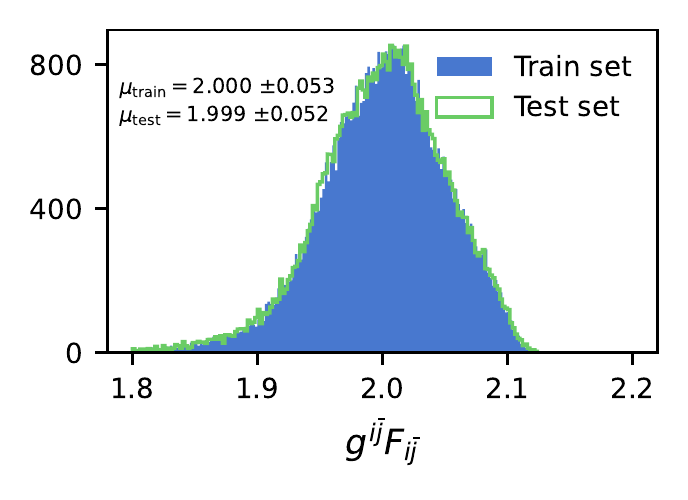}\includegraphics[width=0.3\textwidth]{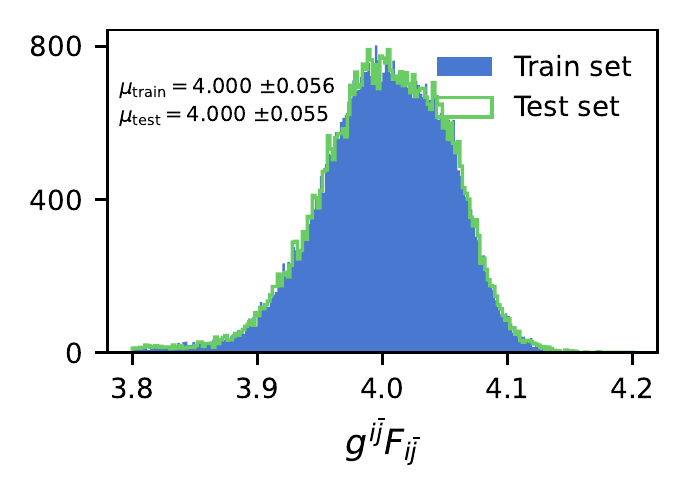}\includegraphics[width=0.3\textwidth]{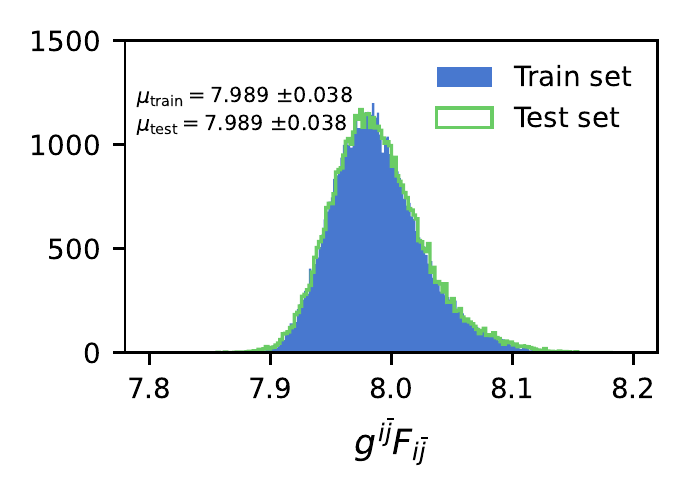}

\caption{Results for line bundle connections on a quintic threefold trained using $\text{Loss}[\boldsymbol{v}]$. The plots show the histogram of $g^{i\bar{j}}F_{i\bar{j}}$ evaluated for points in both the training and test sets, together with the mean and standard deviation. The plots are for networks of depth $D=2,3,4$ which correspond to connections on the line bundles $\mathcal{O}_{X}(2)$, $\mathcal{O}_{X}(4)$ and $\mathcal{O}_{X}(8)$.}
\label{fig:quintic}
\end{figure*}

\Figref{all_elliptic_curve} shows the values of $g^{i\bar j} F_{i \bar j}$ for points on the elliptic curve (\ref{eq:ec}) on the patch $z_0=1$ for $D=2,3,4$ networks, completing \Figref{elliptic_curve} given in the main text. Figures \ref{fig:K3} and \ref{fig:quintic} show the histogram of $g^{i\bar j} F_{i \bar j}$ for line bundle connections for both the K3 and quintic threefold examined in the main text. Each figure displays the histograms for networks of depth $D=2,3,4$, corresponding to computing the connection on $\mathcal{O}_X(2)$, $\mathcal{O}_X(4)$ and $\mathcal{O}_X(8)$. Both the training set and test set are included; there are no signs of overtraining.

\begin{figure*}
\includegraphics[width=0.3\textwidth]{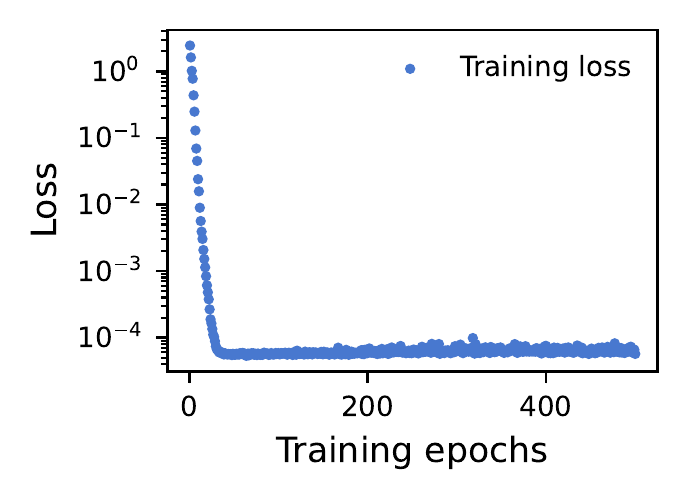}\includegraphics[width=0.3\textwidth]{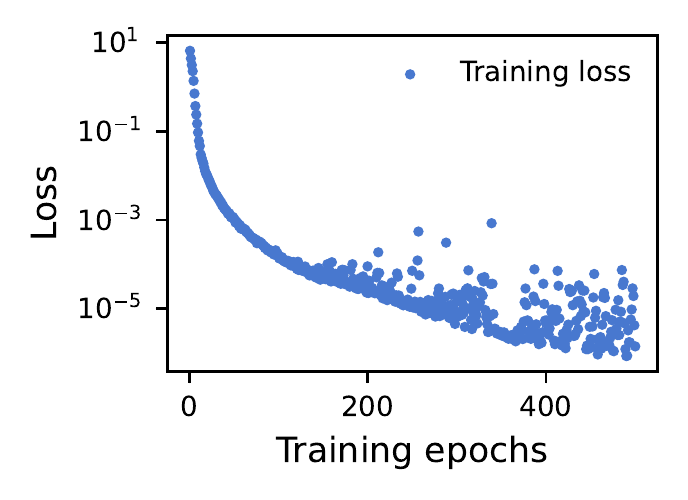}\includegraphics[width=0.3\textwidth]{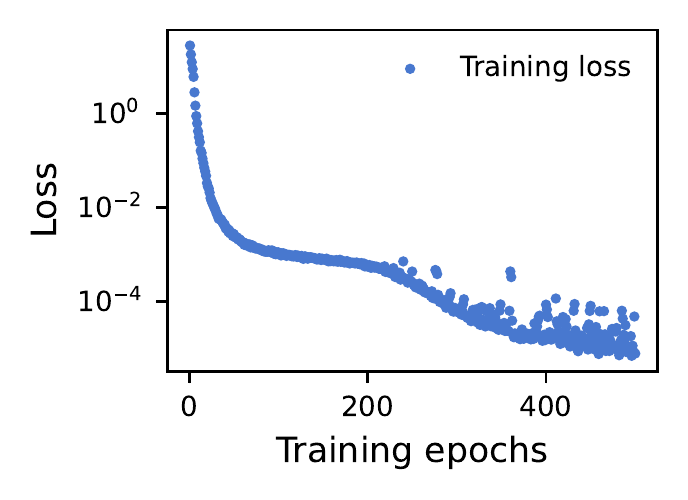}

\caption{Training curves (the value of the loss function evaluated on the training set) as a function of training epoch for line bundle connections on an elliptic curve trained using $\text{Loss}[\boldsymbol{v}]$. The plots are for networks of depth $D=2,3,4$ which correspond to connections on the line bundles $\mathcal{O}_{X}(2)$, $\mathcal{O}_{X}(4)$ and $\mathcal{O}_{X}(8)$.}
\label{fig:ec_training}
\end{figure*}

\begin{figure*}
\includegraphics[width=0.3\textwidth]{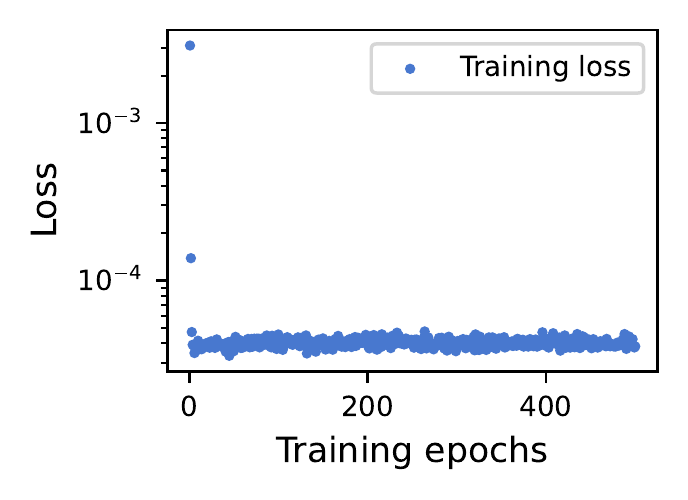}\includegraphics[width=0.3\textwidth]{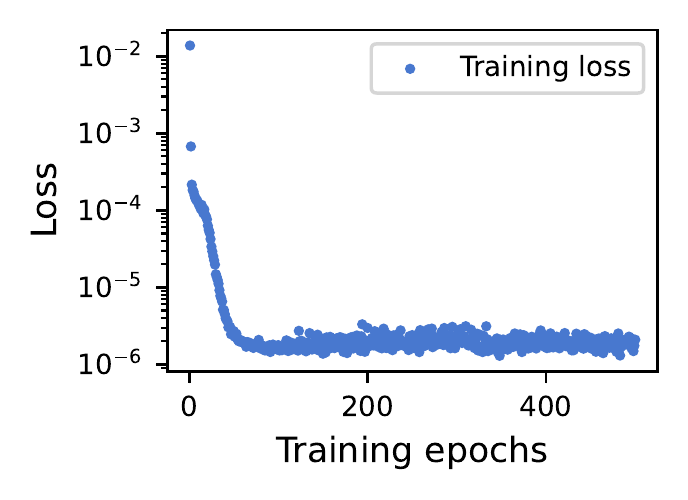}\includegraphics[width=0.3\textwidth]{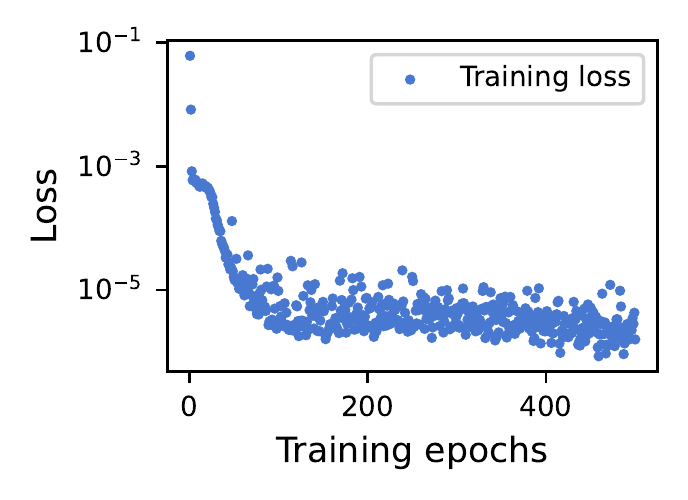}

\caption{Training curves (the value of the loss function evaluated on the training set) as a function of training epoch for line bundle connections on a K3 surface trained using $\text{Loss}[\boldsymbol{v}]$. The plots are for networks of depth $D=2,3,4$ which correspond to connections on the line bundles $\mathcal{O}_{X}(2)$, $\mathcal{O}_{X}(4)$ and $\mathcal{O}_{X}(8)$.}
\label{fig:K3_training}
\end{figure*}

\begin{figure*}
\includegraphics[width=0.3\textwidth]{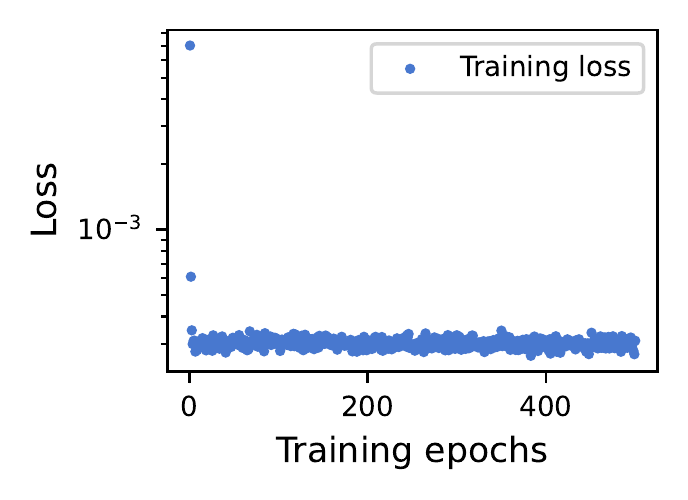}\includegraphics[width=0.3\textwidth]{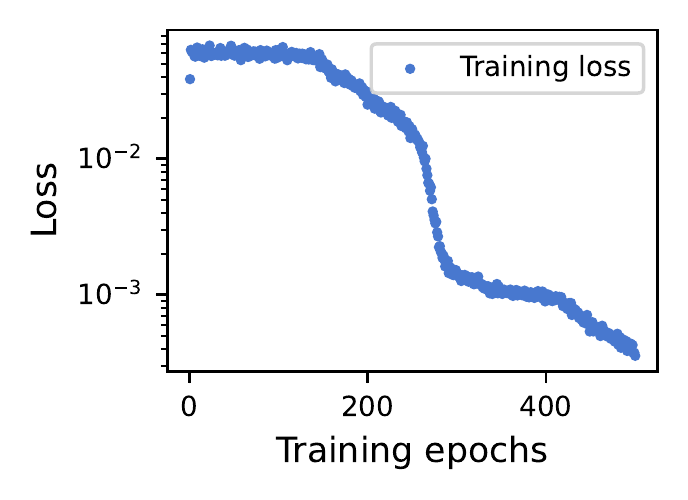}\includegraphics[width=0.3\textwidth]{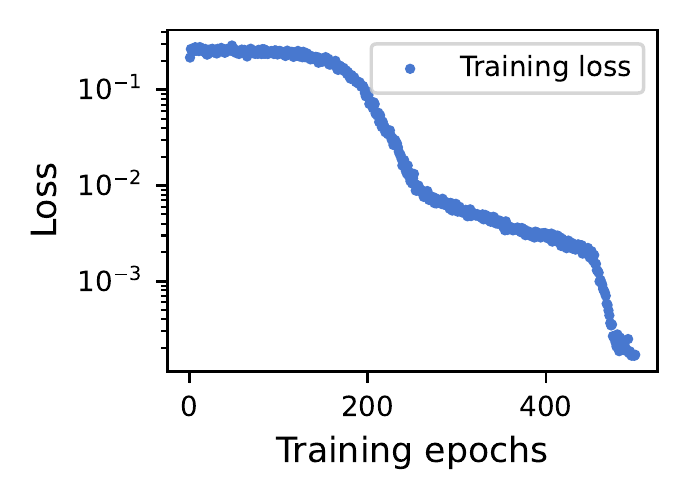}

\caption{Training curves (the value of the loss function evaluated on the training set) as a function of training epoch for line bundle connections on a quintic threefold trained using $\text{Loss}[\boldsymbol{v}]$. The plots are for networks of depth $D=2,3,4$ which correspond to connections on the line bundles $\mathcal{O}_{X}(2)$, $\mathcal{O}_{X}(4)$ and $\mathcal{O}_{X}(8)$.}
\label{fig:quintic_training}
\end{figure*}

Finally, Figures \ref{fig:ec_training}, \ref{fig:K3_training} and \ref{fig:quintic_training} show the training curves (the value of the loss function evaluated on the training set) as a function of training epoch for line bundle connections on the elliptic curve (\ref{eq:ec}), the K3 surface (\ref{eq:K3}), and the quintic threefold (\ref{eq:quintic}), respectively. Each figure displays the training curves for networks of depth $D=2,3,4$, corresponding to computing the connection on $\mathcal{O}_X(2)$, $\mathcal{O}_X(4)$ and $\mathcal{O}_X(8)$. The general pattern that one observes is that the shallow $D=2$ networks reach a minimum very quickly, while the deepest $D=4$ networks show decreasing losses all the way to the 500th epoch.

\end{document}